\documentclass[letterpaper,twocolumn,10pt]{article}
\usepackage{usenix}
\usepackage{etoolbox}
\usepackage{xurlshawn}

\makeatletter
\patchcmd{\maketitle}
  {\@maketitle}
  {\vspace{-0.3in}\@maketitle\vspace{-0.4in}}
  {}
  {}
\makeatother

\usepackage{balance}
\usepackage{url}
\usepackage{color}
\usepackage{caption}
\usepackage{diagbox}
\usepackage{subfigure}
\usepackage{mathptmx}
\usepackage{algorithm,algpseudocode}
\usepackage{amsmath,amssymb}

\usepackage{multirow}
\usepackage{booktabs}
\usepackage{epsfig,endnotes}
\usepackage{enumitem}
\usepackage[font=footnotesize,labelfont=bf, labelsep=period]{caption}
\newcommand{\para}[1]{{\vspace{2pt} \noindent \textbf{#1}
    \hspace{6pt}}}

\newcommand{\shawn}[1]{{\color{black} #1}}

\newcommand{\revise}[1]{{\color{black} #1}}

\newcommand{\etal}{{\em et al.\ }}
\newcommand{\eg}{{\em e.g.,\ }}
\newcommand{\ie}{{\em i.e.,\ }}

\newcommand{\secspace}{\vspace{-0.1in}}

\newcommand{\system}{{\em Glaze}}

\newcommand{\dalleM}{DALL$\cdot$E-m}

\newenvironment{packed_itemize}{
\begin{list}{\labelitemi}{\leftmargin=0.5em}
  \setlength{\itemsep}{1pt}
  \setlength{\parskip}{0pt}
  \setlength{\parsep}{0pt}
  \setlength{\headsep}{0pt}
  \setlength{\topskip}{0pt}
  \setlength{\topmargin}{0pt}
  \setlength{\topsep}{0pt}
  \setlength{\partopsep}{0pt}
}{\end{list}}

\begin{document}

\title{Glaze: Protecting Artists from Style Mimicry by Text-to-Image Models}

\author{Shawn Shan, Jenna Cryan, Emily Wenger, Haitao Zheng, Rana Hanocka, Ben Y. Zhao\\
	{\em Department of Computer Science, University of Chicago}\\
	{\em \{shawnshan, jennacryan, ewillson, htzheng, ranahanocka, ravenben\}@cs.uchicago.edu}}
\maketitle

\begin{abstract}
  Recent text-to-image diffusion models such as MidJourney and Stable
  Diffusion threaten to displace many in the professional artist
  community. In particular, models can learn to mimic the artistic style of
  specific artists after ``fine-tuning'' on samples of their art. In this
  paper, we describe the design, implementation and evaluation of \system{},
  a tool that enables artists to apply ``style cloaks'' to their art before
  sharing online. These cloaks apply barely perceptible perturbations to
  images, and when used as training data, mislead generative models that try
  to mimic a specific artist. In coordination with the professional artist
  community, we deploy user studies to more than 1000 artists, assessing
  their views of AI art, as well as the efficacy of our tool, its usability
  and tolerability of perturbations, and robustness across different
  scenarios and against adaptive countermeasures. Both surveyed artists and
  empirical CLIP-based scores show that even at low perturbation levels ($p$=0.05),
  \system{} is highly successful at disrupting mimicry under normal
  conditions (>92\%) and against adaptive countermeasures (>85\%).

\end{abstract}

\secspace
\section{Introduction}
\label{sec:intro}

\begin{figure}[t]
  \centering
  \includegraphics[width=1\columnwidth]{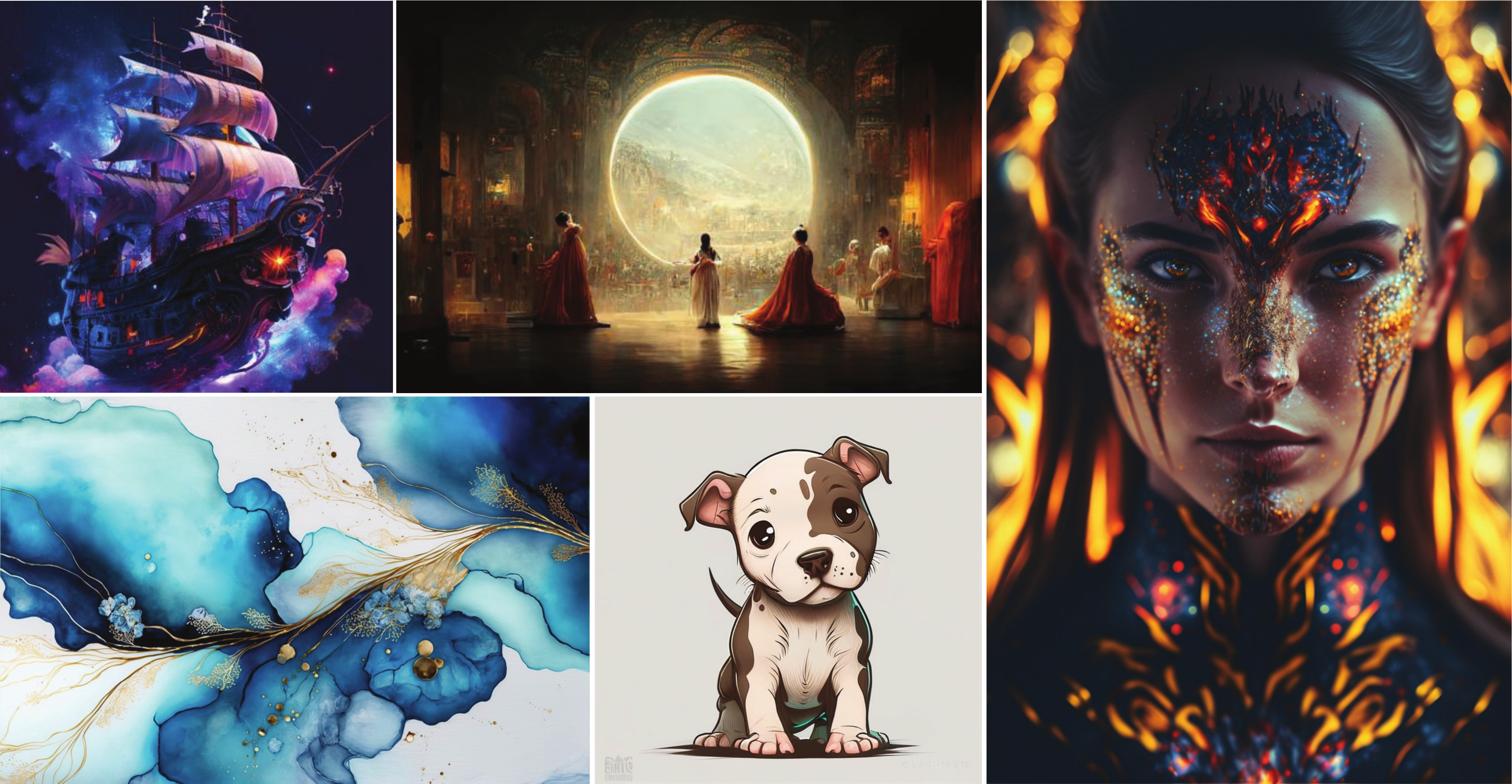}
  \vspace{-0.25in}
  \caption{Sample AI-generated art pieces from the Midjourney
    community showcase \cite{mid-top-artistname,winaward}. }
  \label{fig:aiart}
\end{figure}

It is not an exaggeration to say that the arrival of text-to-image generator
models has transformed, perhaps upended, the art industry. By sending simple
text prompts like ``A picture of a corgi on the moon'' to diffusion models
such as StableDiffusion or MidJourney, anyone can generate incredibly
detailed, high resolution artwork that previously required many hours of work
by professional artists. AI-art such as those in Figure~\ref{fig:aiart} have
won awards at established art conventions~\cite{winaward}, served as
cover images for magazines~\cite{c-ai-cover}, and used to
illustrate children's books~\cite{children-book} and video
games~\cite{aigame}. More powerful models continue to
arrive~\cite{stable2-1,mid-4,novelai-update}, catalyzed by VC
funding~\cite{sd-funding,scenario-vc,other1-funding}, technical research
breakthroughs~\cite{kawar2022imagic,chen2022re,meng2022distillation,li2022scaling,balaji2022ediffi},
and powered at their core by continuous training on a large volume of human-made
art scraped from online art repositories such as ArtStation, Pinterest and DeviantArt. 

Only months after their arrival, these models are rapidly growing in users
and platforms. In September 2022, MidJourney reported over 2.7 million
users and 275K AI art images generated {\em each
  day}~\cite{midjourneystat}. Beyond simple prompts, many have taken the open
sourced StableDiffusion model, and ``fine-tuned'' it on additional samples
from specific artists, allowing them to generate AI art that {\em mimics} the
specific artistic styles of that artist~\cite{aiart-greg}. In fact, entire
platforms have sprung up where home users are posting and sharing their own
customized diffusion models that specialize on mimicking specific artists, likeness of
celebrities, and NSFW themes~\cite{civitai}. 


Beyond open questions of copyrights~\cite{AIcopyright}, ethics~\cite{AIethics,AIethics2}, and
consent~\cite{AIconsent,AIconsent2,AIconsent3}, it is clear that these AI models have had significant negative
impacts on independent artists. For the estimated hundreds of thousands of
independent artists across the globe, most work on commissions, and attract
customers by advertising and promoting samples of their artwork
online. First, professional artists undergo years of training to develop
their individual artistic styles. A model that mimics this style profits from
that training without compensating the artist, effectively ending their
ability to earn a living. Second, as synthetic art mimicry continues to grow
for popular artists, they displace original art in search results, further
disrupting the artist's ability to advertise and promote work to potential
customers~\cite{aiart-forbes,aiart-greg}. Finally, these mimicry attacks are
demoralizing art students training to be future artists. Art students see
their future careers replaced by AI models even if they can successfully find
and develop their own artistic styles~\cite{studentsquit}.

Today, all of these consequences have indeed occurred in the span of a few
months. Art students are quitting the field; AI models that mimic specific
artists are uploaded and shared for free; and professional artists are losing
their livelihoods to models mimicking their unique styles. Artists are
fighting back via lawsuits~\cite{ailawsuit,class-action}, online boycotts and
petitions~\cite{aiprotest}, but legal and regulatory action can take years,
and are difficult to enforce internationally. Thus most artists are faced a
choice to 1) do nothing, or 2) stop sharing samples of their art online to
avoid training models, and in doing so cripple their main way to advertise
and promote their work to customers.

In this paper, we present the design, implementation and evaluation of a
technical alternative to protect artists against style mimicry by
text-to-image diffusion models. 
We present Glaze, a system that allows an artist to apply carefully
computed perturbations to their art, such that diffusion models will learn
significantly altered versions of their style, and be ineffective in future
attempts at style mimicry. We worked closely with members of the professional
artist community to develop Glaze, and conduct multiple user studies with
1,156 participants from the artist community to evaluate its efficacy,
usability, and robustness against a variety of active countermeasures.

Intuitively, \system{} works by taking a piece of artwork, and computing a
minimal perturbation (a ``style cloak'') which, when applied, shifts the artwork's
representation in the generator model's feature space towards a chosen target
art style. Training on multiple cloaked images teaches the generator model
to shift the artistic style it associates with the artist, leading to mimicry
art that fails to match the artist's true style.

{Our work makes several key contributions:}
\vspace{-0.1in}
\begin{packed_itemize}
\item We engage with top professional artists and the broader community, and
  conduct user studies to understand their views and concerns towards AI art
  and the impact on their careers and community.
\item We propose \system{}, a system that protects artists from style mimicry
  by adding minimal perturbations to their artwork to mislead AI models to
  generate art different from the targeted artist. 92\% of surveyed artists
  find the perturbations small enough not to disrupt the value of their art. 
\item Surveyed artists find that \system{} successfully disrupts
  style mimicry by AI models on protected artwork. 93\% of 
  artists rate the protection is successful under a variety of settings,
  including tests against real-world mimicry platforms.
\item In challenging scenarios where an artist has already posted significant
  artworks online, we show \system{} protection remains high. 87.2\% of
  surveyed artists rate the protection as successful when an artist is only
  able to cloak 1/4 of their online art (75\% of art is uncloaked).
\item We evaluate \system{} and show that it is robust (protection success
  > 85\%) to a variety of adaptive countermeasures.
\item We discuss \system{} deployment and post-deployment
  experiences, including countermeasures in the wild.  
\end{packed_itemize}

\para{Ethics.} Our user study was reviewed and approved by our institutional
review board (IRB). All art samples used in experiments were used with
explicit consent by their respective artists. All user study participants
were compensated for their time, although many refused payment. 

\secspace
\section{Background: AI Art and Style Mimicry}
\label{sec:motivation}

In this section, we provide critical context in the form of
basic background on current AI art models and style mimicry.

\secspace
\subsection{Text-to-Image Generation}

\begin{figure}[t]
  \centering
  \includegraphics[width=1\columnwidth]{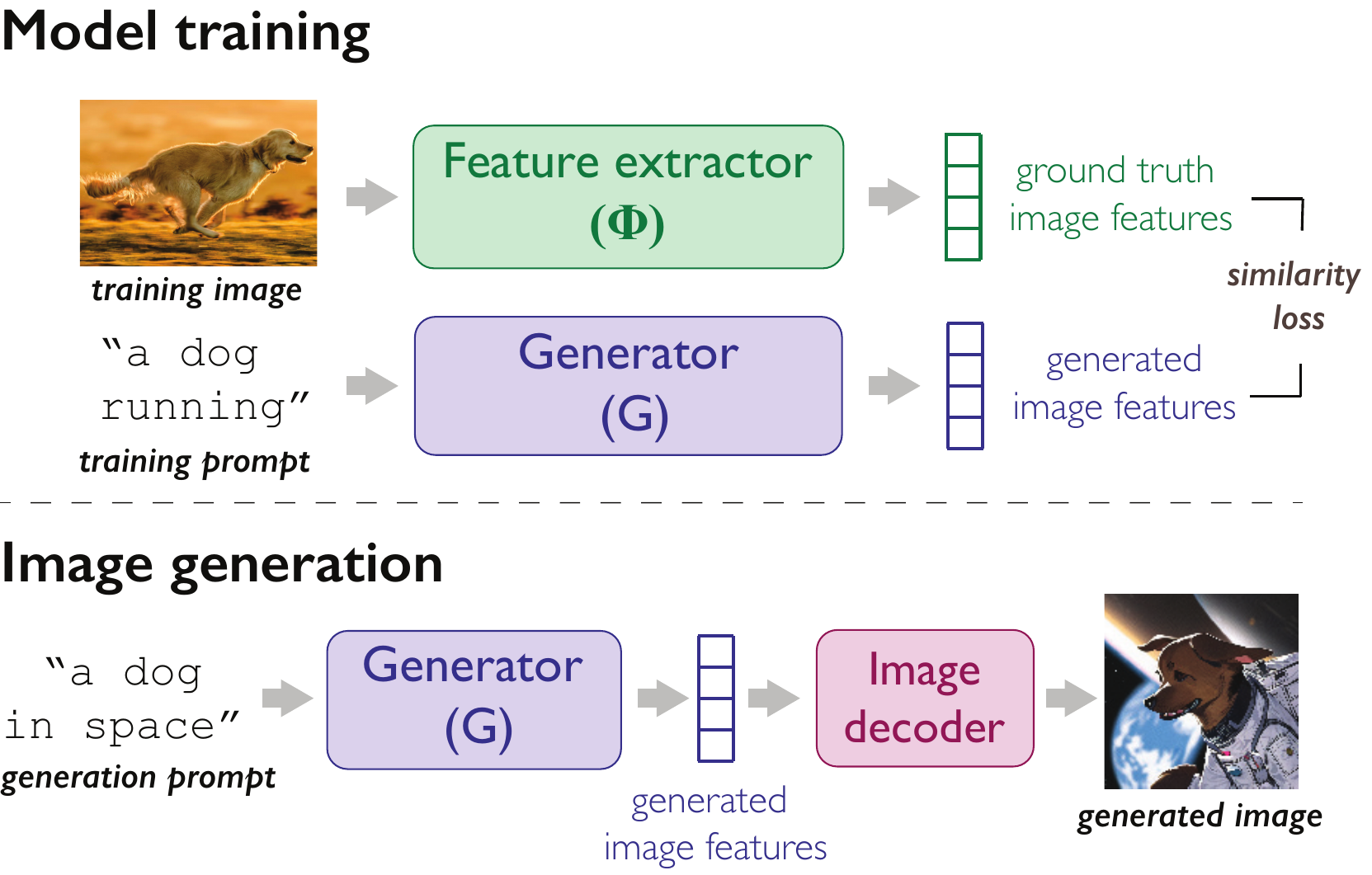}
  \vspace{-0.25in}
  \caption{High level model architecture of text-to-image models. }
  \label{fig:diffusion-arch}
\end{figure}

Since Text-to-image generation was first proposed in
2015~\cite{mansimov2015generating}, a stream of research has proposed newer
model architectures and training methods enabling generation of
higher-quality
images~\cite{radford2015unsupervised,zhang2017stackgan,xu2018attngan,li2019object,zhu2019dm}.  The high level design of
recent models used for AI art
generation~\cite{rombach2022high,ramesh2021zero,d-mini} is shown in
Figure~\ref{fig:diffusion-arch}. During training, the model takes in an image
$x$ and uses a feature extractor $\Phi$ to extract its features, producing
$\Phi(x)$. Simultaneously, a conditional image generator $G$ takes in a
corresponding text caption ($s$) and outputs a predicted feature vector
$G(s)$. Then the parameters of $G$ are optimized so the text feature vector
$G(s)$ matches the image feature vector $\Phi(x)$. At generation time, a user
gives $G$ a text prompt $s_0$, and $G$ outputs an image feature vector
$G(s_0)$. A decoder $D$ then decodes $G(s_0)$ to produce the final generated
image. 

Compared to earlier models based on generative adversarial networks
(GANs) or variational autoencoders
(VAE)~\cite{radford2015unsupervised,zhu2019dm,tao2022df}, more recent
models~\cite{rombach2022high,ramesh2022hierarchical}
leveraging \textit{diffusion} models
produce significantly higher quality images. Feature extractor ($\Phi$) is
used to reduce the dimensionality of the input image to facilitate the
generation process. The extractor $\Phi$ and decoder $D$ are often a pair of
variational autoencoder (VAE)~\cite{rombach2022high,ramesh2021zero}, \ie
extractor (encoder) extracts image features and decoder map features back to
images.

\para{Training Data Sources. } The training datasets of these models
typically contain image/ALT text pairs scraped from the Internet. They
are extremely large, e.g. LAION~\cite{schuhmann2022laion} contains 5 billion
images collected from 3 billion webpages.  

These datasets are subject to minimal curation and governance. Data
collectors typically only filter out data with extremely short or incorrect text captions
(based on an automated text/image alignment metric~\cite{schuhmann2022laion}). 
Since copyrighted images are not filtered~\cite{schuhmann2022laion}, these
datasets are rife with private, sensitive content, including copyrighted
artworks.

\secspace
\subsection{Style Mimicry}
\label{sec:mimicry-back}

\begin{figure}[t]
  \centering
  \includegraphics[width=0.95\columnwidth]{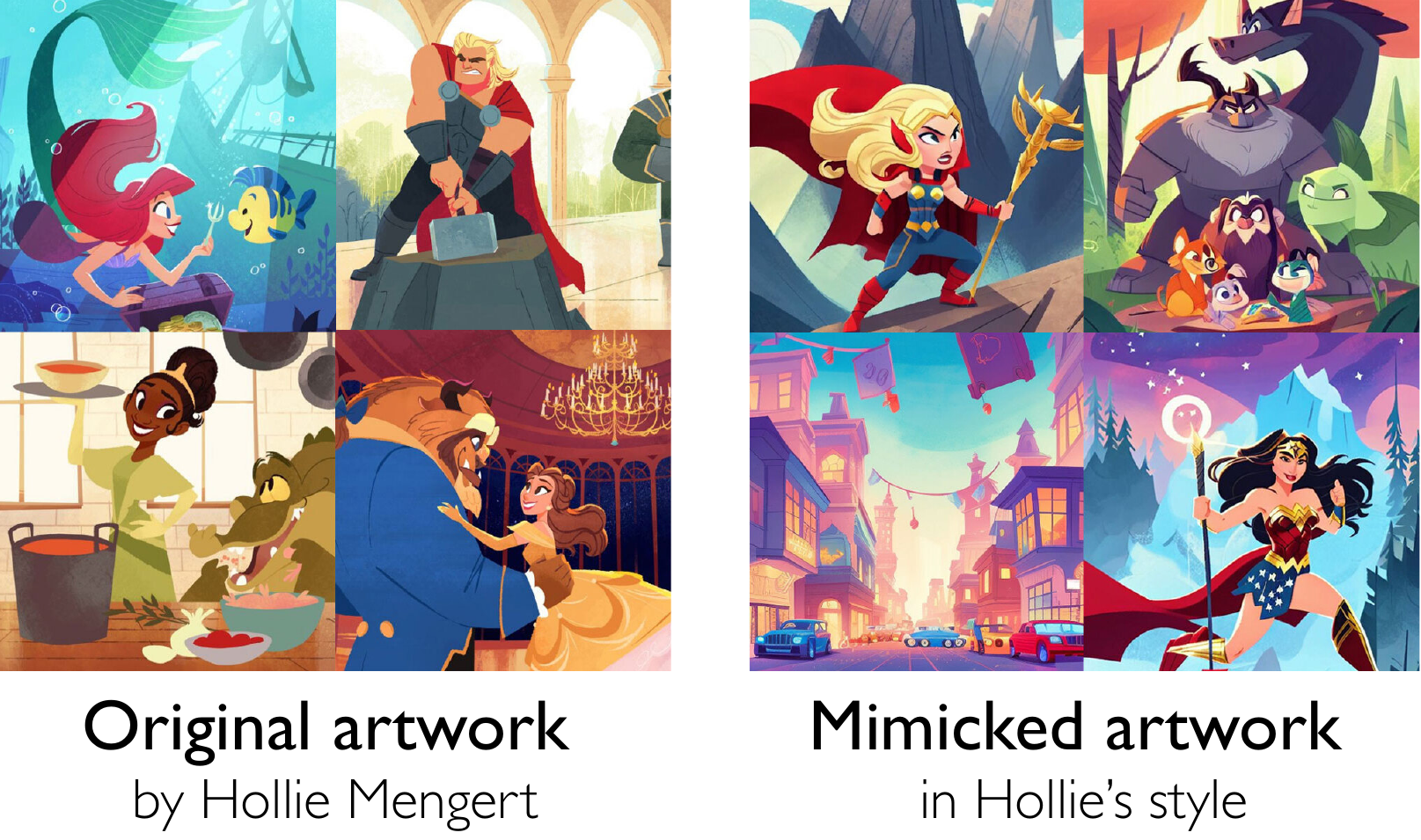}
  \vspace{-0.1in}
  \caption{Real-world incident of AI plagiarizing the style of artist Hollie Mengert~\cite{hollie-steal}. {\bf Left}: original artwork by Hollie Mengert. {\bf Right}: plagiarized artwork generated by a model trained to mimic Hollie's style. }
  \label{fig:hollie-mimic}
\end{figure}

In a {\em style mimicry} attack, a bad actor uses an AI art model to create
art in a particular artist's style without their consent. 
More than 67\% of 
art pieces showcased on a popular AI-art-sharing website leverage style
mimicry~\cite{mid-top-artistname}.

\para{Style mimicry techniques.} Today, a ``mimic'' can easily copy the style of
a victim artist with only an open-source text-to-image model and a few
samples of artwork from the artist. A naive mimicry attack directly queries a
generic text-to-image model using the name of the victim artist. For example,
the prompt ``a painting in the style of Greg Rutkowski'' would cause the
model to generate images in the style of Polish artist Greg Rutkowski. This
is because many of Rutkowski's artworks appear in training datasets of these
generic models labeled with his name.

Naive mimicry can succeed when the artist is well-known and
has a significant amount of art online, but fail on other artists. In more recent
mimicry attacks, a mimic {\em fine-tunes} a
generic text-to-image model on samples of a target artist's work
(as few as $20$ unique pieces) downloaded from online sources. This calibrates the model to
the victim artist's style, identifying important features related to the
victim style and associating these regions in the feature space with the
victim artist's name~\cite{ruiz2022dreambooth,gal2022image}. This enables
style mimicry with impressive accuracy.  The entire fine-tuning process takes
less than 20 minutes on a low-end consumer GPU\footnote{It takes an average
  of 18.3 minutes on a GTX 1080 GPU}.

\para{Real-work mimicry incidents. }
The first well-known incident of mimicry was when a Reddit user stole
American artist Hollie Mengert's style and open-sourced the style-specific
model on Reddit~\cite{hollie-steal}. Figure~\ref{fig:hollie-mimic} has a
side-by-side comparison of Hollie's original artwork and plagiarized artwork
generated via style mimicry. Later, famous cartoonist Sarah Andersen reported
that AI art models can mimic her cartoon drawings~\cite{sarah-andersen}, and
other similar incidents abound~\cite{lensa-steal,sam-steal}.

Several companies~\cite{aigame} have even hosted style mimicry
as a service, allowing users to upload a few art pieces painted by victim
artists and producing new art in the victim styles. CivitAI~\cite{civitai}
built a large online marketplace where people share their customized stable
diffusion models, fine-tuned on certain artwork.  

\secspace
\section{Collaborating with Artists}
\label{sec:artists}
Next, we explain our collaborative relationship with professional artists,
and its significant impact on our key evaluation metrics in this paper. We
also summarize key results from our first user study on views of AI art and
mimicry by members of the artist community.

Artists have spoken out against style mimicry in numerous venues, focusing
particularly on how it violates their intellectual property rights and
threatens their
livelihoods~\cite{guardian-artical,artical-1,artical-2, noai-protest}. Others
have taken direct action. The Concept Art Association raised over \$200K to
fight AI art, and filed a class action lawsuit in the US
against AI art companies~\cite{class-action}. In November 2022, artists
organized a large protest against ArtStation~\cite{noai-protest}, the large
digital art sharing platform that allowed users to post AI artwork without
identification. Anti-AI images flooded the site for several weeks, until
ArtStation banned the protest images~\cite{noai-result}. \revise{Recently, the
Writers Guild of America (WGA) went on strike demanding contractual changes
to ban generative AI~\cite{AIstrike}.}

Members of the professional art community reached out to us in Sept 2022. We
joined online town halls and meetings alongside hundreds of professionals,
including Emmy winners and artists at major film studios. After learning
more, we began an active collaboration with multiple professional artists,
including award-winning artist Karla Ortiz, who leads efforts defending
artists and is lead plaintiff in the class action suit.
The artists helped this project in multiple ways, by 1) sharing experiences
about specific ways AI-art has impacted them and their colleagues; 2) sharing
domain knowledge about what is acceptable to artists in terms of
perturbations on their art; and 3) helping to widely disseminate our user study to
members of their professional organizations, including the Concept Art
Association and the Animation Guild (TAG839).

\para{Evaluation via Direct Feedback from Artists.} Our goal is to help artists
disrupt AI models trying to mimic their artistic style, without adversely
impacting their own artwork. Because ``success'' in this context is highly
subjective (``Did this AI-art successfully mimic Karla's painting style?''),
we believe the only reliable evaluation metric is direct feedback by
professional artists themselves. Therefore, wherever possible, the evaluation
of \system{} is done via detailed user studies engaging members of the
professional artist community, augmented by an empirical score we develop based on
genre prediction using CLIP models.

We deployed two user studies during the course of this project (see
Table~\ref{tab:study-details}). Both are IRB-approved by our institution.  Both
draw participants from professional artists informed via their social circles
and professional networks. The first (Survey 1, \S\ref{sec:user-study},
\S\ref{sec:cloaking-results}), asked participants
about their broad views of AI style mimicry, and then presented them with a
number of inputs and outputs of our tool, and asked them to give ratings
corresponding to key metrics we wanted to evaluate. We select a subset 
of participants from the first study to participate in a 
longer and more in-depth study (Survey 2) where 
they were asked to evaluate the performance of \system{} in 
additional settings (\S\ref{sec:cloaking-results}, \S\ref{sec:robust-eval}, 
\S\ref{sec:counter}, and Appendix~\ref{sec:appendix}).

\begin{table}[t]
  \centering
  \resizebox{0.49\textwidth}{!}{
  \centering
    \begin{tabular}{ccl}
    \hline
    \textbf{Survey} & \textbf{\begin{tabular}[c]{@{}c@{}} \# of artists\end{tabular}} & \multicolumn{1}{c}{\textbf{Content}} \\ \hline
    Survey 1 & 1156 & \begin{tabular}[c]{@{}l@{}} 1) Broad views of AI art
                        and style mimicry(\S\ref{sec:user-study}) \\
                        2) Glaze's usability, i.e. acceptable levels of cloaking (\S\ref{sec:cloaking-results}) \\
                        3) Glaze performance in disrupting style mimicry (\S\ref{sec:cloaking-results}) \end{tabular} \\ \hline
    
\begin{tabular}[c]{@{}c@{}}
    Survey 2\\(Extension to Survey 1)
    \end{tabular}
    & 151 & \begin{tabular}[c]{@{}l@{}}
    1) Additional performance tests (\S\ref{sec:cloaking-results}) \\
    2) Robustness to advanced scenarios (\S\ref{sec:robust-eval}) \\ and countermeasures (\S\ref{sec:counter}) \\ 
    3) Additional system evaluation (Appendix~\ref{sec:appendix}) \end{tabular} \\ \hline
    \end{tabular}
  }
  \vspace{-0.1in}
  \caption{Information on our user studies: the number of artist participants
    and where we report the results of the studies. We sent Survey 2 to
    some specific participants from survey 1 who volunteered to participate in a
    followup study.}
  \label{tab:study-details}
\end{table}

\secspace
\subsection{Artists' Opinions on Style Mimicry}
\label{sec:user-study}

While we expected artists to view style mimicry negatively, we wanted to
better understand how much individual artists understood this topic and how
many perceived it as a threat. Here we describe results from Survey 1 to
gather perceptions of the potential impact of AI art on existing artists.

\para{Survey Design.} Our survey consisted of both multiple choice and free
response questions to understand how well people understand the concept of AI
art, and how well the models successfully imitate the style of artists.
Additionally, we asked artists about the extent to which they anticipate the
emergence of AI art to impact their artistic activities, such as posting
their art online and their job security.  A handful of professional artists
helped disseminate our survey to their respective artist community groups.
Overall, we collected responses from 1,207 participants, consisting primarily
of professional artists (both full-time (46\%) and part-time/freelancer
(50\%)) and some non-artist members of the art community who felt invested in
the impact of AI art (4\%). Of the participants who consider themselves
artists, their experience varied: <1 year (13\%), 1-5 years (49\%), 5-10
years (19\%), 10+ years (19\%).  Participants' primary art style varied
widely, including: animation, concept art, abstract, anime, game art, digital
2D/3D, illustration, character artwork, storyboarding, traditional
painting/drawing, graphic design, and others.

\para{Key Results.} Our study found that 91\% of the artists have read about
AI art extensively, and either know of or worry about their art being used to
train the models. Artists expect AI mimicry to have a significant impact on artist
community: $97\%$ artists state it will decrease some artists' job security;
$88\%$ artists state it will discourage new students from studying art; and
$70\%$ artists state it will diminish creativity. ``Junior positions will
become extinct,'' as stated by one participant.

Many artists (> 89\% artists) have already or plan to take actions because of
AI mimicry. Over $95\%$ of artists post their artwork online. Out of these
artists, $53\%$ of them anticipate reducing or removing their online artwork,
if they haven't already. Out of these artists, $55\%$ of them believe
reducing their online presence will significantly impact their careers. One
participant stated ``AI art has unmotivated myself from uploading more art
and made me think about all the years I spent learning art.'' $78\%$ of
artists anticipate AI mimicry would impact their job security, and this
percentage increases to $94\%$ for the job security of newer
artists. Further, $24\%$ of artists believe AI art has \textit{already}
impacted their job security, and an additional $53\%$ expect to be affected
within the next 3 years. Over $51\%$ of artists expressed interest in
proactive measures, such as personally joining class action lawsuits against
AI companies.  

Professional artists thought AI mimicry was very successful at mimicking the
style of specific artists.  We showed the artists examples of original
artwork from $23$ artists, and the artwork generated by a model
attempting to mimic their styles (detailed mimicry setup in
\S\ref{sec:eval-cloak}).  $77\%$ of artists found the AI model
\textit{successfully} or \textit{very successfully} mimic the styles of
victim artists, with one stating ``it's shocking how well AI can mimic the
original artwork.''  Additionally, $19\%$ of participants thought the AI
mimicry is somewhat successful, leaving only $< 5\%$ of artists rating the
mimicry as unsuccessful.  Several artists also pointed out that, as artists,
upon close inspection they could spot differences between the AI art and
originals, but were skeptical the general public would notice them.

A significant concern of most participants, surprisingly, is not just the
existence of AI art, but rather scraping of existing artworks without
permission or compensation.  As one participant stated: ``If artists are paid
to have their pieces be used and asked permission, and if people had to pay
to use that AI software with those pieces in it, I would have no problem.''
However, without consent to use their artwork to train the models, ``it's
incredibly disrespectful to the artist to have their work `eaten' by a
machine [after] many years to grow our skills and develop our styles.''
\begin{figure*}[t]
  \centering
  \includegraphics[width=0.85\linewidth]{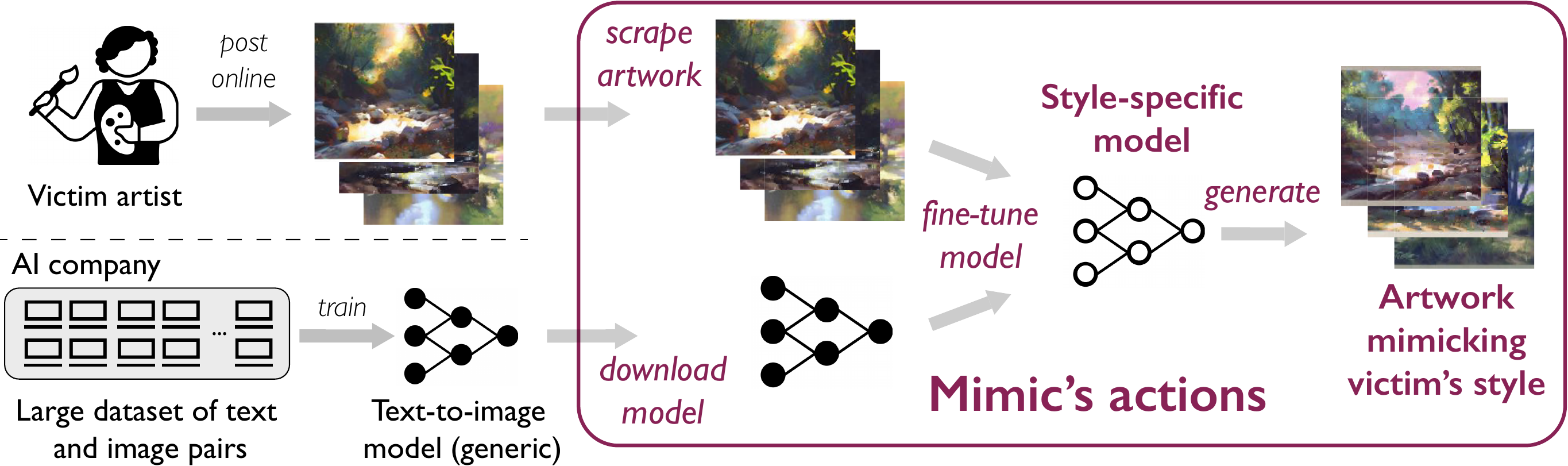}
  \caption{High level overview of the mimicry attack scenario. The mimic scrapes copyrighted artwork from the victim artist and uses these to fine-tune a pre-trained, generic text-to-image model. \shawn{The generic model is trained and open-sourced by an AI company. }The mimic then uses the fine-tuned model to generate artwork in the style of the victim artist. }
  \label{fig:scenario}
\end{figure*}

\secspace
\section{Preliminaries}
\label{sec:cloak}

We propose \system{}, a tool that protects artists against AI
style mimicry. An artist uses \system{} to add small digital perturbations
(``cloak'') to images of their own art before sharing online
(Figure~\ref{fig:cloaking}). A text-to-image model that trains on cloaked
images of artwork will learn an incorrect representation of the artist's style in
feature space \ie the model's internal understanding of artistic styles. When
asked to generate art pieces in victim's style, the model will fail to mimic
the style of the victim, and instead output art pieces in a recognizably 
different style. 

Here, we first introduce the threat model, then discuss existing alternatives
to the AI style mimicry problem. We present the intuition behind \system{} and detailed
design in \S\ref{sec:design}.

\secspace
\subsection{Threat Model}

Here we state assumptions for both the artists protecting their own art
and the users training models to replicate their artistic style. We
refer to these AI art model trainers as ``mimics.''

\begin{figure}[t]
  \centering
  \includegraphics[width=0.9\columnwidth]{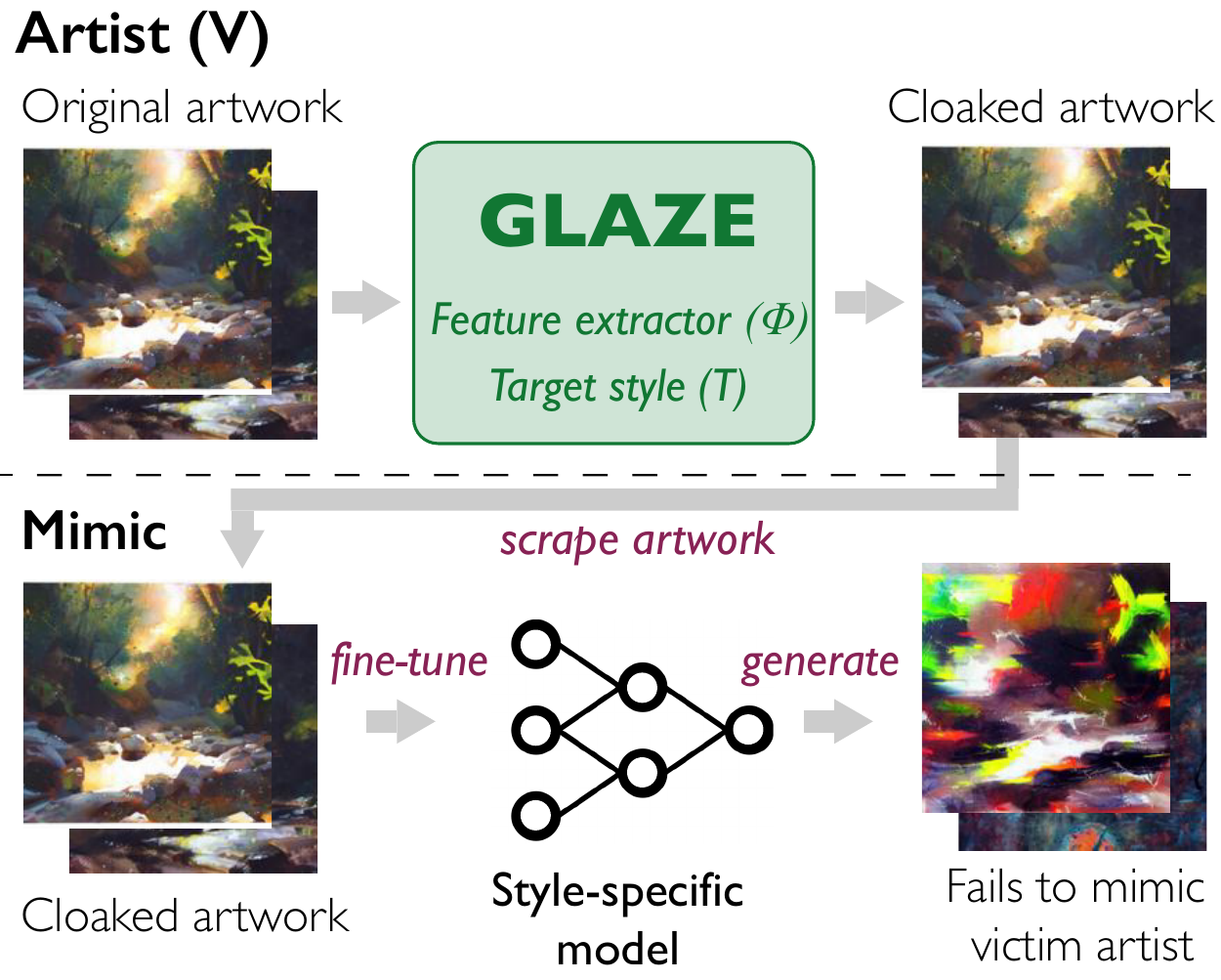}
  \caption{Overview of \system{}, a system that protects victim artists from AI style mimicry by cloaking their online artwork. ({\bf Top}) An artist $V$ applies the cloaking algorithm (uses a feature extractor $\Phi$ and a target style $T$) to generate cloaked versions of $V$'s art pieces. Each cloak is a small perturbation unnoticeable to human eye. ({\bf Bottom}) A mimic scrapes the cloaked art pieces from online and uses them to fine-tune a model to mimic $V$'s style. When prompted to generate artwork in the style of $V$, mimic's model will generate artwork in the target style $T$, rather than $V$'s true style. }
  \label{fig:cloaking}
\end{figure}

\para{Artists.} Artists want to share and promote their artwork online
without allowing mimics to train models that replicate their art styles.
Sharing art online enables artists to sell their work and attract
commissioned work, fueling their livelihoods~(\S\ref{sec:artists}). Artists
protect themselves by adding imperceptible perturbations to their artwork
before sharing them as shown in Figure~\ref{fig:cloaking}. The goal of the
\system{} cloak is to disrupt the style mimicry process, while only
introducing minimal perturbation on images of the artwork.

We assume the artists have access to moderate computing resources (\eg a
laptop) and add perturbation to images of their artwork locally before
posting online. We also assume artists have access to some public feature
extractor (\eg open-source models such as Stable Diffusion). We begin with
assumption that artists use the same feature extractor as mimics (large
majority of mimics use the open-source Stable Diffusion model). We later
relax this assumption.

\para{Mimics. } The mimic's goal is to train a text-to-image model that
generates \emph{high-quality} art pieces of any subject in the \emph{victim's
  style}. A mimic could be a well-funded AI company, \eg Stability AI or
OpenAI, or an individual interested in the style of victim artist. We assume
the mimic has:

\vspace{-0.1cm}
\begin{packed_itemize}
\item access to the weights of generic text-to-image models well-trained on large datasets;
\item access to art pieces from the target artist;
\item significant computational power. 
\end{packed_itemize}

\vspace{-0.1cm}
We assume the attack scenario where the mimic 
fine-tunes its model on images of the artist's artwork (as shown in
Figure~\ref{fig:scenario}). This is stronger than the naive mimic attack
without fine tuning. 
Finally, we assume the mimic is aware of our
protection tool and can deploy adaptive countermeasures (\S\ref{sec:counter}).  

\secspace
\subsection{Potential Alternatives and Challenges}
\label{sec:challenge}

A number of related prior works target protection against invasive and
unauthorized facial recognition models. They proposed ``image cloaking'' as a
tool to prevent a user's images from being used to train a facial recognition
model of
them~\cite{shan2020fawkes,cherepanova2021lowkey,chandrasekaran2020face,evtimov2020foggysight,wenger2021sok}. They
share a similar high level approach, by using optimized perturbations that
cause cloaked images to have drastically different feature representations
from original user images. 
It is possible to
adapt existing cloaking-based systems to protect artists against AI style
mimicry. Protection system would compute a cloak on each artwork in order
to perturb its feature space representation to be different from its
unperturbed representation. This can succeed if the cloak significantly shifts
the artwork's feature representation, making resulting models generate
dramatically different artwork.

We found that in practice, however, existing solutions are unable to
introduce large-enough feature space shifts to achieve the desired
protection. This is due to the properties of feature spaces in text-to-image
models. Face recognition models {\em classify identities}, so their feature
spaces mainly represent identity-related information. On the other hand,
text-to-image models {\em reconstruct original images from extracted
  features}, so their feature spaces retain more information about the
original image (objects, locations, color, style, etc.). Thus, producing the
same shift in feature representation in a text-to-image model is much harder
(requires more perturbation budget) than in a classification model. This
observation is validated by prior work showing that adversarial perturbations
are much less effective at attacking generative
models~\cite{kos2018adversarial,gondim2018adversarial,tabacof2016adversarial}. Specifically,
\cite{kos2018adversarial,tabacof2016adversarial} found that adversarial
attack methods that are effective at attacking classifiers are significantly
less effective at attacking autoencoders. We empirically confirm that
existing cloaking methods cannot prevent AI mimicry (\S\ref{app:prior-work}
in Appendix). We show that Fawkes~\cite{shan2020fawkes} and
LowKey~\cite{cherepanova2021lowkey} perform poorly in this setting, even when
artists add highly visible cloaks to their artwork.

For generative models, concurrent work~\cite{madry-defense}
proposes PhotoGuard, a method to cloak images to prevent
unauthorized image edits (inpainting) on cloaked images. Similar to existing
cloaking systems, PhotoGuard tries to indiscriminately minimize all
information contained in an image (\ie the norm of the feature vector) to
prevent models from editing the image. Thus, it is also not effective at
mimicry prevention.

\para{Design Challenges. } The main reason that existing cloaking methods
fail to prevent AI mimicry is because they indiscriminately shift all
features in an image, wasting the cloak perturbation budget on shifting
unnecessary features (\eg object shape, location, etc.). Protecting artist's
style requires only {\em shifting features related to the artistic style of
  victim}. This can be achieved if a text-to-image model learns to draw objects
similar to those drawn by the victim artist {\em as long as the model cannot
  mimic the artist's unique style}. Thus, optimal protection from mimicry requires
concentrating the cloak on style-specific features.  

Unfortunately, identifying and separating out these style-specific features
is difficult. Even assuming the existence of interpretability methods that
perfectly explain the feature space of a text-to-image model, there is no
clear way to mathematically define and calculate ``artistic styles.'' In all
likelihood, any definition would change across different styles. For example,
``impressionist'' likely correlates more strongly with color features,
whereas ``cubism'' correlates with shape features. Even across multiple
art pieces in the same style, the style may manifest differently.  

\begin{figure}[t]
  \centering
  \includegraphics[width=1\columnwidth]{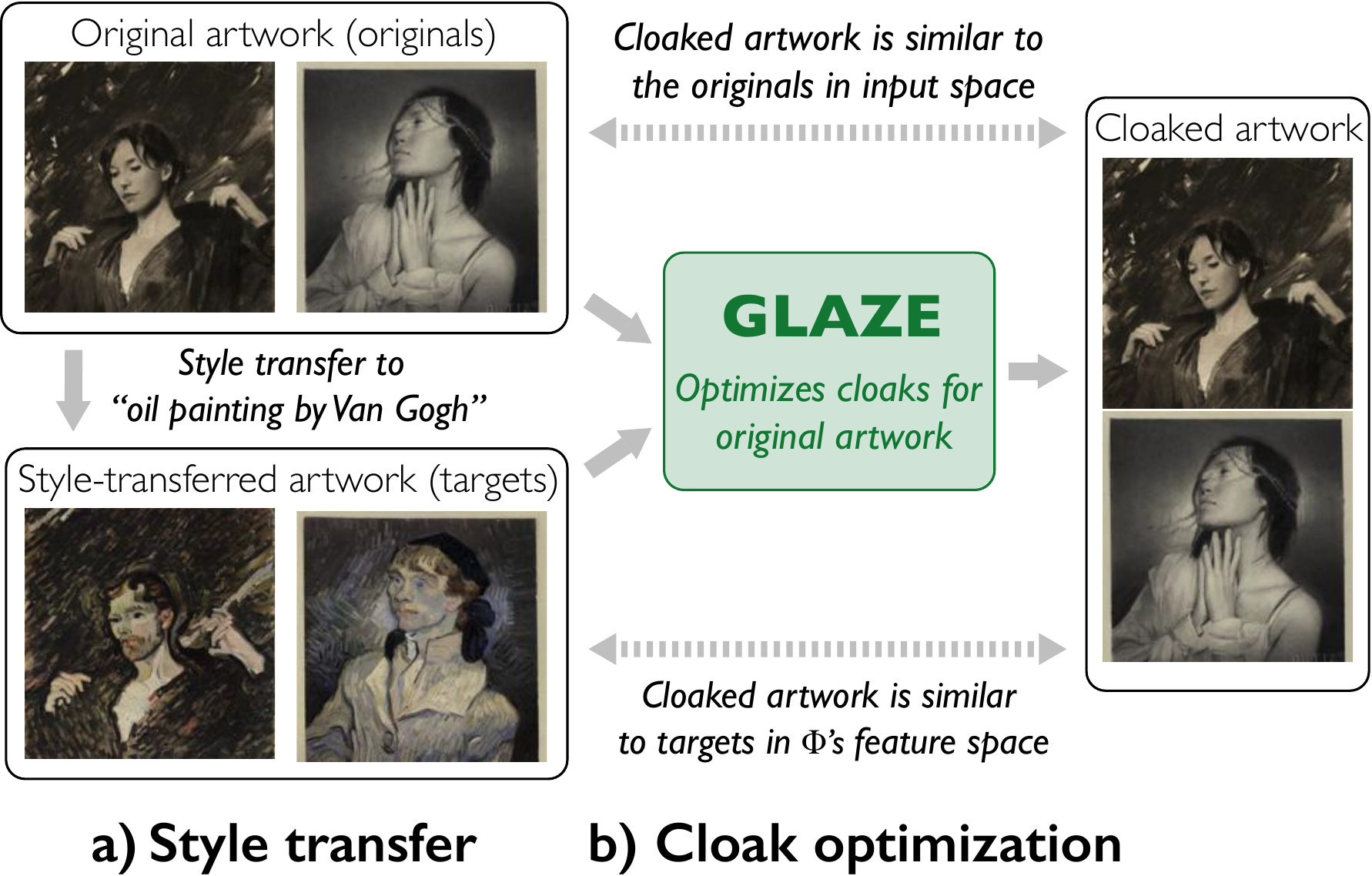}
  \caption{High level overview of how \system{} perturbs the style-specific features of the artwork. {\bf a)} \system{} style transfers the original artwork to a different style, which changes its style but leaves other features unaltered. {\bf b)} \system{} optimizes a cloak that makes the artwork's features representation match that of the style-transferred art, while constraining the amount of visible changes to the artwork.  }
  \label{fig:cloak-intuition}
  \vspace{-0.3cm}
\end{figure}

\begin{figure}[t]
  \centering
  \includegraphics[width=1\columnwidth]{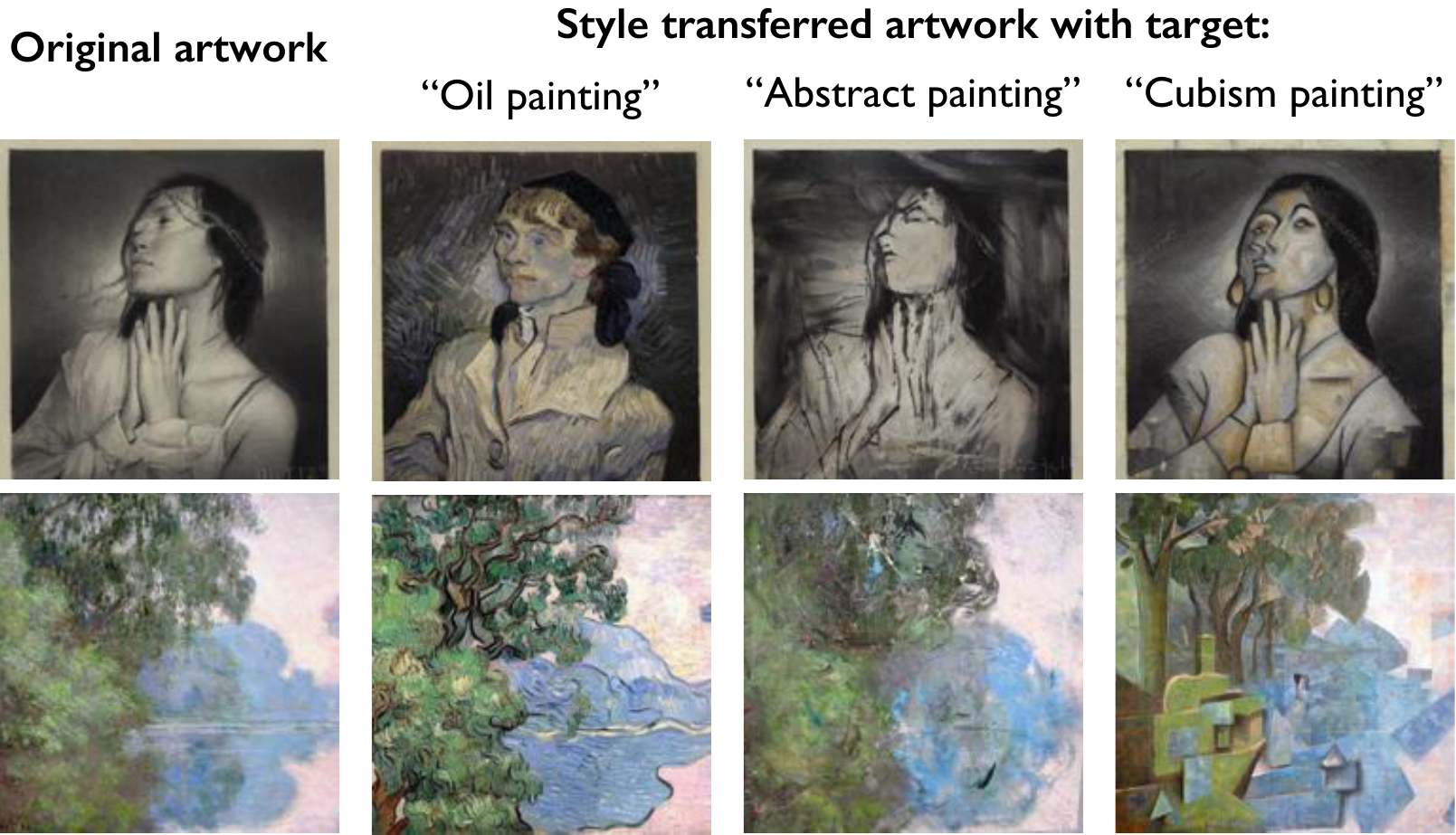}
  \caption{Example style-transferred artwork with different target styles. }
  \label{fig:transfer-examples}
\end{figure}

\secspace
\section{Disrupting Style Mimicry with Glaze} 
\label{sec:design}

In this section, we introduce \system{}, its design intuition followed by the
detailed algorithm. 

\secspace
\subsection{Design Intuition}
\label{sec:intuition-cloak}

Our key intuition is to identify and isolate \textit{style-specific features}
of an artist's original artwork, \ie the set of image features that
correspond to artistic style. Then \system{} computes cloaks while focusing
the perturbation budget on these style-specific features to maximize impact
on stylistic features.

As discussed, identifying and calculating style-specific features in model's
feature space is difficult due to the poor interpretability of model features
and how art style manifests differently across artworks. We overcome these two
challenges by designing a style-dependent and artwork-dependent method that
operates at image space. Given an artwork, we leverage ``style transfer,'' an
end-to-end computer vision technique, to modify and isolate its style
components. ``Style transfer'' transforms an image into a new image with a
different style (\eg from impressionist style to cubist style) while keeping
other aspects of the image similar (\eg subject matter and location).

We leverage style transfer in our protection technique as follows. Given an
original artwork from the victim artist, we apply style-transfer to produce a
similar piece of art with a different style, \eg in style of ``an oil
painting by Van Gogh'' (Figure~\ref{fig:cloak-intuition} a). The new
version has similar content to the original, but its style mirrors that
of Van Gogh. We show more style-transfer examples with different target styles in
Figure~\ref{fig:transfer-examples}. Now, we can use the style-transferred
artwork as projection target to guide the perturbation computation. This
perturbs the original artwork's style-specific features towards that of
the style-transferred version. We do this by optimizing a cloak that, when
added to the original artwork, makes its feature representation similar to
the style-transferred image. Since the content is identical between the pair
of images, cloak optimization will focus its perturbation budget on style
features. 

\secspace
\subsection{Computing Style Cloaks} 

Using this approach, we compute style cloaks to disrupt style mimicry as
follows. Given an artwork ($x$), we use an existing feature extractor to
compute the style-transferred version of $x$ into target style $T$: $\Omega(x, T)$.
We then compute a style cloak $\delta_x$, such that $\delta_x$ moves $x$'s
style-specific feature representation to match that of $\Omega(x, T)$ while
minimizing visual impact. The cloak generation optimization is:

\secspace
\begin{eqnarray}
   &\min\limits_{\delta_x} Dist\left( \Phi(x + \delta_x), \Phi (\Omega(x, T))\right),  \label{eq:cloakopt}\\
  & \text{subject to } \; |\delta_x|< p, \nonumber
\end{eqnarray} 

\noindent where $\Phi$ is a generic image feature extractor commonly used in
text-to-image generation tasks, $Dist(.)$ computes the distance of two
feature representations, $|\delta_x|$ measures the perceptual perturbation
caused by cloaking, and $p$ is the perceptual perturbation budget. 

As discussed in \S\ref{sec:intuition-cloak}, the use of the style-transferred
image $\Omega(x, T)$ guides the cloak optimization in Eq~(\ref{eq:cloakopt})
to focus on changing style-specific image features. To maximize cloak
efficacy, the target style $T$ should be dissimilar from artist's original
style in the feature space. We discuss our heuristic for selecting target styles in
\S\ref{sec:design}.

\secspace
\subsection{Detailed System Design}
\label{sec:design-details}

Now we present the detailed design of \system{}. Given a victim
artist $V$, \system{} takes as input the set of $V$'s artwork to be shared
online $X_V$, an image feature extractor $\Phi$, a style-transfer model
$\Omega$, and perturbation budget $p$. Note that in many cases, a single
model (e.g. Stable Diffusion) provides both $\Phi$ and $\Omega$.

\para{Step 1: Choose Target Style.}  The selected target style $T$
should be sufficiently different from $V$'s style in model feature space to
maximize chances of disrupting style mimicry. For example, Fauvism and
Impressionism are distinct art styles that often look visually similar to the
untrained eye. Image of an impressionist painting style cloaked to Fauvism
might not produce a visually discernible effect on model-generated
paintings. Note that an artist can maximize their ability to avoid mimicry if
they consistently style cloak all their artwork towards the same target $T$.

For a new user, \system{} uses the following algorithm to randomly select $T$ from a set
of candidate styles reasonably different from $V$'s style. The algorithm
first inspects a public dataset of artists, each with a specific style (\eg
Monet, Van Gogh, Picasso). For each candidate target artist/style, it selects
a few images in that style and calculates their feature space centroid using
$\Phi$. It also computes $V$'s centroid in $\Phi$ using $V$'s artwork. Then,
it locates the set of candidate styles whose centroid distance to $V$'s
centroid is between the $50$ to $75$ percentile of all candidates. Finally,
it randomly selects $T$ from the candidate set.

\para{Step 2: Style transfer. } \system{} then leverages a pre-trained
style-transfer model $\Omega$~\cite{rombach2022high} to generate the
style-transferred artwork for optimization. Given each art piece $x \in
X_V$ and target style $T$, it style transfers $x$ to target style $T$ to
produce style-transferred image $\Omega(x, T)$.  

\para{Step 3: Compute cloak perturbation.} Then, \system{} computes the
cloak perturbation, $\delta_x$ for $x$, following the optimization defined
by eq. (\ref{eq:cloakopt}), subject to $|\delta_x| < p$. Our implementation
uses LPIPS (Learned Perceptual Image Patch
Similarity)~\cite{zhang2018unreasonable} to bound the perturbation. Different
from the $L_p$ distance used in previous
work~\cite{carlini2017towards,kurakin2016adversarial,sabour2015adversarial},
LPIPS has gained popularity as a measure of user-perceived image
distortion~\cite{cherepanova2021lowkey,laidlaw2020perceptual,rony2021augmented}. Bounding
cloak generation with this metric ensures that cloaked versions of images are
visually similar to the originals. We apply the \textit{penalty
  method}~\cite{nocedal2006numerical} to solve the optimization in
eq.(\ref{eq:cloakopt}) as follows:  

\vspace{-0.05in}
\begin{equation} \label{eq:optdetail} \vspace{-0.03in} 
 \underset{\delta_x}{\text{min }}||\Phi(\Omega(x, T)), \Phi(x + \delta_x) ||_2^2 + \alpha \cdot max(LPIPS(\delta_x)-p, 0) 
\end{equation}

\noindent where $\alpha$ controls the impact of the input perturbation. $L_2$
distance is used to calculate feature space distance.

\para{Upload artwork online.} Finally, the artist posts the cloaked artwork
online. For artists already with a large online presence, they can cloak and
re-upload artwork on their online portfolio.
While updating online images is not always possible, 
\system{} can be effective even when the mimic's model has significant amount
of uncloaked art (\S\ref{sec:robust-eval}).

\secspace
\subsection{On the Efficacy of Style Cloaks}
\label{sec:cloak-effect}

\system's style cloaks work by shifting feature representation of artwork in
the generator model. But how much shift do we need in order to have a
noticeable impact on mimicked art?

Two reasons suggest that even small shifts in style will have a meaningful
impact in disrupting style mimicry. First, generative models used for style
mimicry have {\em continuous} output spaces, \ie any shift in image feature
representation results in changes in the generated image.  Because generative
models are trained to interpolate their continuous feature
spaces~\cite{white2016sampling,upchurch2017deep}, any shift in the model's
representation of art style results in a new style, a ``blend'' between the
artist and the chosen target style.  
Second, mimicked artwork must achieve reasonable quality and similarity in
style to the artist to be useful. Small shifts in the style space often
produce incoherent blends of conflicting styles that are enough to disrupt
style mimicry, \eg thick oil brushstrokes of Van Gogh's style mixed into a realism portrait.

These two factors contribute to \system{}'s success in more challenging
scenarios (\S\ref{sec:robust-eval}), and its robustness against
countermeasures (e.g. adversarial training) that succeed against cloaking
tools for facial recognition (\S\ref{sec:counter}).

\section{Evaluation}
\label{sec:eval-cloak}

In this section, we evaluate \system's efficacy in protecting artists from
style mimicry. We first describe the datasets, models, and experimental
configurations used in our tests. Then we present the results of \system's
protection in a variety of settings. Due to \system's highly visual nature,
we evaluate its performance using both direct visual assessment by
\textbf{human artists} in a user study, and \textbf{automated metrics} (see
\S\ref{sec:metrics} for details).

\para{Summary of results.} Over $93\%$ of artists surveyed believe \system{}
effectively protects artists' styles from AI style mimicry
attacks. Protection efficacy remains high in challenging settings, like when
the mimic has access to unprotected artwork. \system{} also achieves high
protection performance against a real-world mimicry-as-a-service platform. Of
our $1156$ artist participants, over $92\%$ found the perturbations
introduced by cloaking small enough not to disrupt the value of their art,
and over $88\%$ would like to use \system{} to protect their own artwork from
mimicry attacks.

\secspace
\subsection{Experiment Setup}
\label{sec:cloak-setup}

\para{Mimicry dataset. } We evaluate \system's performance in protecting the styles of the following two groups of artists: 

\vspace{-0.2cm}
\begin{packed_itemize}
\item {\em Current artists}: $4$ professional artists let us use their
  artwork in our experiments. These artists have different styles and
  backgrounds (\eg full-time/freelancers, watercolor painters/digital
  artists, well-known/independent). Each provided us with between $26$ to
  $34$ \textit{private} original art pieces for our experiments. We use
  perceptual hashing~\cite{ke2004efficient} to verify that none of these are
  included in existing public datasets used to train generic text-to-image
  models (e.g.~\cite{schuhmann2022laion,changpinyo2021conceptual}).  

\item {\em Historical artists}: We also evaluate \system{}'s protection on
  $195$ historical artists (\eg van Gogh, Monet) from the WikiArt
  dataset~\cite{saleh2015large}. The WikiArt dataset contains 42,129 art
  pieces from $195$ artists. Each art piece is labeled with its genre (\eg
  impressionism, cubism). We randomly sampled $30$ art pieces from each
  artist to use in style mimicry attacks. Generic text-to-image models found
  online have been trained on some artwork from these artists. Using this art
  simulates a more challenging scenario in which a famous artist attempts
  to disrupt a model that already understands their style.
\end{packed_itemize}
\vspace{-0.2cm}

\para{Mimicry attack setup. } We recreate the strongest-possible mimicry
attack scenario, based on techniques used in real-world mimicry
incidents~\cite{ruiz2022dreambooth,sam-steal,hollie-steal},
that works as follows. First, we take art pieces from the victim artist $V$
and generate a text caption for each piece using an image captioning
model~\cite{luo2022vc}. \revise{The pretrained image captioning model generates a short 
sentence to describe the image. We found that this model can correctly caption protected images (examples in Figure~\ref{fig:data-examples}), likely because \system{} focuses on perturbing style features while the captioning models focus on image content.} Then, we append the artist's name to each caption,
\eg ``mountain range \textit{by Vincent van Gogh}''. Finally, we fine-tune a pre-trained generic text-to-image model
(details below) on the caption/image pairs. 

We use $80\%$ of the art pieces from the victim artists to fine-tune models
that mimic each artist's style, reserving the rest for testing. We fine-tune
for $3000$ optimization steps, which we find achieves the best mimicry
performance (Figure~\ref{fig:success-iter} in Appendix). We then use the
fine-tuned, style-specific model to generate mimicked artwork in style of
each victim artist. We query the model using the generated captions (which
include $V$'s name) from the held-out test artwork set. We generate $5$
pieces of mimicked art for each text caption using different random seeds and
compare these to the real victim art pieces with this caption. Additional
details on training and generation parameters, as well as its sensitivity to
random seed selection and the number of training art pieces are in Appendix~\ref{app:mimicry}.  


\para{Text-to-image models.} We use two state-of-the-art, public, generic text-to-image models in our experiments: 

\vspace{-0.2cm}
\begin{packed_itemize}
\item \textit{Stable Diffusion (SD)}: Stable Diffusion is a popular and
  high-performing open-source text-to-image model~\cite{stable2-1},trained
  on 11.5 million images from the LAION dataset~\cite{schuhmann2022laion}. SD
  training takes over 277 GPU months (on A100 GPU) and costs around
  \$600K~\cite{stable2-1}. SD uses diffusion methods to generate images and
  achieves state-of-the-art performance on several
  benchmarks~\cite{rombach2022high}. Viewed as one of the best open-source
  models, SD has powered many recent developments in text-to-image
  generation~\cite{blender-plugin,novelai-update,gimp,aigame}. We
  use SD version 2.1 in the paper~\cite{stable2-1}, the most up-to-date
  version as of December 2022.  

\item \textit{DALL$\cdot$E-mega (\dalleM)}: \dalleM-mega, an updated version
  of the more well-known \dalleM-mini, is an open-source model based on
  OpenAI's \dalleM~1~\cite{ramesh2021zero}. The model leverages a VAE for
  image generation and is trained on 17 million images from three different
  datasets~\cite{sharma-etal-2018-conceptual,changpinyo2021conceptual,thomee2016yfcc100m}. Training
  takes 2 months on 256 TPUs~\cite{mini-training}. While \dalleM~ performs
  worse than diffusion-based models like SD, we use it to evaluate how
  \system~generalizes to different model architectures.  
\end{packed_itemize}

\vspace{-0.2cm}

\para{\system~configuration. } We generate cloaks for each of victim $V$'s
art pieces following the methodology of \S\ref{sec:design-details}. First, we
use the target selection algorithm to select a target style $T$. We choose
from a set of $1119$ candidate target styles, collected by querying the
WikiArt dataset with artist and genre names, \eg ``Impressionism painting by
Monet''~\footnote{One artist may paint in multiple styles, resulting in
  multiple candidate target styles from a single artist.}. We then style
transfer each victim art piece into the target style leveraging the style
transfer functionality of stable diffusion model (stable diffusion model has
both text-to-image and style transfer functionality). \revise{A style transfer model takes
in an original image and a target prompt as input. Leveraging a similar diffusion process, the model
modifies the original image to a style similar to that described in the target prompt. More information on style transfer can be found in ~\cite{saharia2022palette}}. Finally, we optimize a cloak for each art piece
using Eq.~\ref{eq:optdetail} by running the Adam optimizer for $500$
steps. \revise{We benchmark \system{}'s runtime on artwork with resolution ranging 
from $512$ to $6000$ pixels, using SD's feature extractor (ViT model with 83 million parameters). } It takes an average of $1.2$ mins on Titan RTX GPU and $7.3$ mins on a
single Intel i7 CPU to generate a cloak for a single piece of art. 

In our initial experiments, we assume \system{} generates cloaks using the
same image feature extractor as the mimic (e.g. SD's or \dalleM's feature
extractor). We relax this assumption and evaluate
\system{}'s performance when artists and mimics use different feature
extractors in \S\ref{sec:robust-eval}.

\secspace
\subsection{Evaluation Metrics} 
\label{sec:metrics}

We evaluate our protection performance using both visual assessment and feedback from
human artists, and a scalable metric. Here, we describe the setup of our
evaluation study and define the exact metrics used for evaluation.  

\para{Artist-rated protection success rate (Artist-rated PSR): } The user
studies ask artists to rate the performance of \system. We generate a dataset
of mimicry attacks on $13$ victim artists (the $4$ current artists and $9$
randomly chosen historical artists) across $23$ protection scenarios
(including ones in \S\ref{sec:counter}). For each participant, we randomly
select a set of mimicry attacks out of these $13 \times 23$ settings and ask
them to evaluate protection success.  For each mimicry attempt, we show
participants $4$ mimicked art pieces and $4$ original art pieces from the
victim artist. \revise{Using original art pieces as an indicator of the human
artist's style,} we ask participants to consider the mimicked art, and rate the success of 
\system{}'s
protection on a 5-level Likert scale (ranging from ``not successful at all''
to ``very successful''). Each mimicry attempt is evaluated by at least $10$
participants. We define \textit{artist-rated PSR} as the percent of
participants who rated \system{}'s protection as ``successful'' or ``very
successful.''  Our user studies primarily focus on artists, as they would be
most affected by this technology. We found though, that not all current
artists despise AI art, and some view it as a new avenue for a different form
of artistry.

\para{CLIP-based genre shift: } We define a new metric based on
CLIP~\cite{radford2021learning}, using the intuition that \system{} succeeds
if the mimicked art has been impacted enough by \system{} to be classified
into a \textit{different art genre} from the artist's original artwork. We
leverage CLIP model's ability to classify art images into art genres. Given a
set of mimicked art targeting an artist $V$, we define \textit{CLIP-based
  genre shift rate} as the percentage of mimicked art whose top 3 predicted
genres do not contain $V$'s original genre. A higher genre shift rate means
more mimicked art belongs to a different genre from the victim artist, and thus
means more successful protection.

To calculate the genre shift we use a set of $27$ historical genres from
WikiArt dataset and $13$ digital art genres~\cite{digital-styles} as the
candidate output labels. In Appendix~\ref{app:clip}, we show that a
pre-trained CLIP model is able to achieve high genre classification
performance. We report the average CLIP-based genre shift for all 199 victim
artists across all mimicked artworks.

We use CLIP-based genre shift as a supplemental metric to evaluate \system{}
because it is only able to detect style changes at the granularity of art
genres.
However, mimicry attacks also fail when
\system{} causes the mimicked artwork quality to be very low, something that 
CLIP cannot measure. Measuring the quality of generated image has been a
challenging and ongoing research problem in computer
vision~\cite{kynkaanniemi2022role,blau2018perception,karras2020training}.

\begin{figure*}[t]
  \centering
  \includegraphics[width=0.9\linewidth]{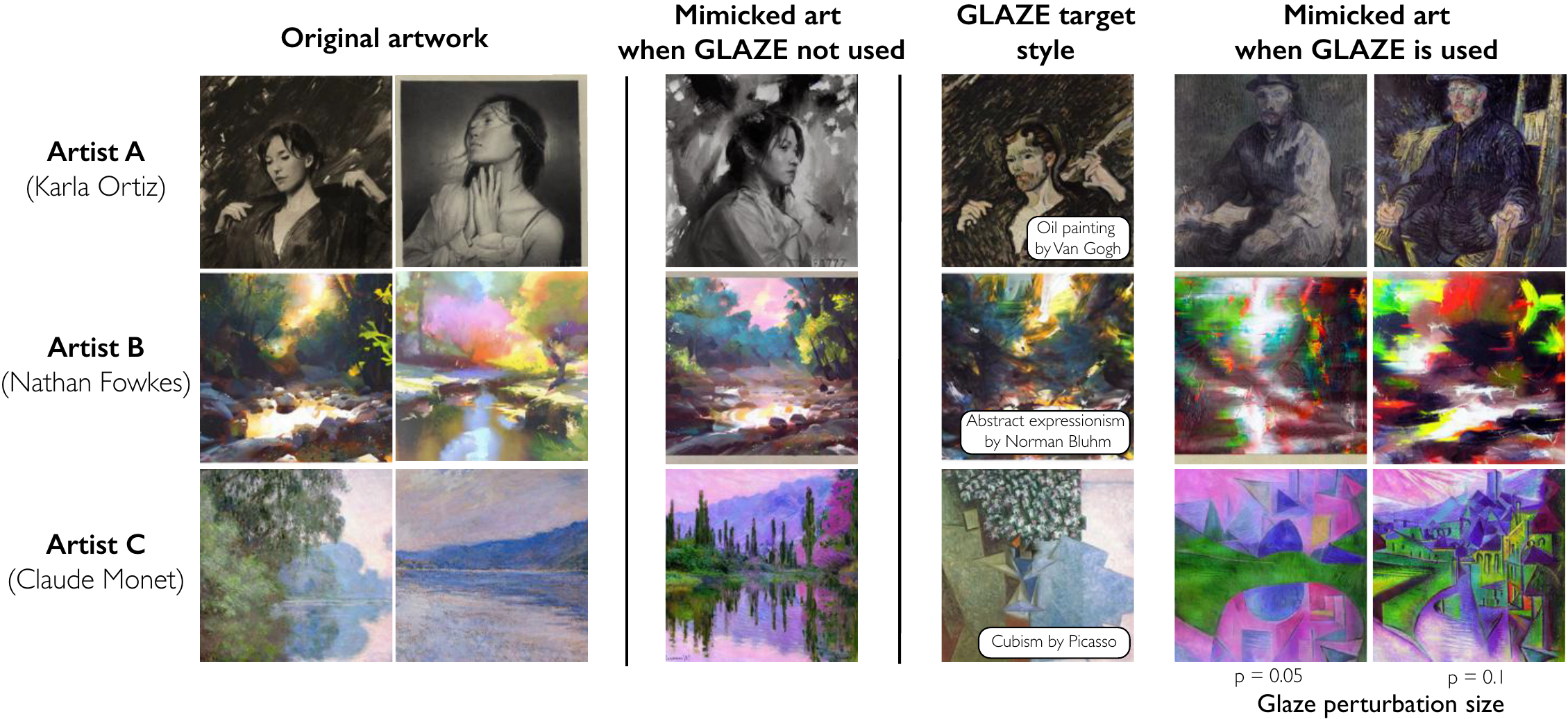}
  \vspace{-0.1in}
  \caption{Example \system{} protection results for three artists. {\bf
      Columns 1-2}: artist's original artwork; {\bf column 3}: mimicked
    artwork when artist does not use protection; {\bf column 4}:
    style-transferred artwork (original artwork in column 1 is the source)
    used for cloak optimization and the name of target style; {\bf column
      5-6}: mimicked artwork when artist uses cloaking protection with
    perturbation budget $p=0.05$ or $p=0.1$ respectively. All mimicry
    examples here use SD-based models.
  } 
  \label{fig:core-res}
\end{figure*}

\begin{table}[t]
  \centering
  \resizebox{0.5\textwidth}{!}{
  \centering
\begin{tabular}{cccccc}
\toprule
\multirow{2}{*}{\textbf{\begin{tabular}[c]{@{}c@{}} \\ Generic \\ model\end{tabular}}} & \multirow{2}{*}{\textbf{\begin{tabular}[c]{@{}c@{}}\\ Artist \\ dataset\end{tabular}}} & \multicolumn{2}{c}{\textbf{w/o \system{}}} & \multicolumn{2}{c}{\textbf{w/ \system{} (p=0.05)}} \\ \cmidrule{3-6} 
 &  & \begin{tabular}[c]{@{}c@{}}Artist-rated \\ PSR\end{tabular} & \begin{tabular}[c]{@{}c@{}}CLIP-based \\ genre shift\end{tabular} & \begin{tabular}[c]{@{}c@{}}Artist-rated \\ PSR\end{tabular} & \begin{tabular}[c]{@{}c@{}}CLIP-based \\ genre shift\end{tabular} \\ \midrule
\multirow{2}{*}{SD} & Current & $4.6 \pm 0.3\%$ & $2.4 \pm 0.2\%$ & $94.3 \pm 0.8\%$ & $96.4 + 0.5\%$ \\
 & Historical & $4.2 \pm 0.2\%$ & $1.3 \pm 0.2\%$ & $93.3 + 0.6\%$ & $96.0 + 0.3\%$ \\ \midrule
\multirow{2}{*}{\dalleM~} & Current & $31.9 \pm 3.5\%$ & $6.4 \pm 0.8\%$ & $97.4 \pm 0.2\%$ & $97.4 + 0.3\%$ \\
 & Historical & $29.8 \pm 2.4\%$ & $5.8 \pm 0.6\%$ & $96.8 \pm 0.3\%$ & $97.1 + 0.2\%$ \\ \bottomrule
\end{tabular}
  }
  \vspace{-0.1in}
  \caption{\system{} has a high protection success rate, as measured by
    artists and CLIP, against style mimicry attacks. We compare protection
    success when artists do not use \system{} vs. when they do (with
    perturbation budget 0.05). }
  \label{tab:psr-core-table}
  \vspace{-0.3cm}
\end{table}

\begin{figure*}[t]
  \centering
  \begin{minipage}{0.32\textwidth}
  \centering
  \includegraphics[width=1\columnwidth]{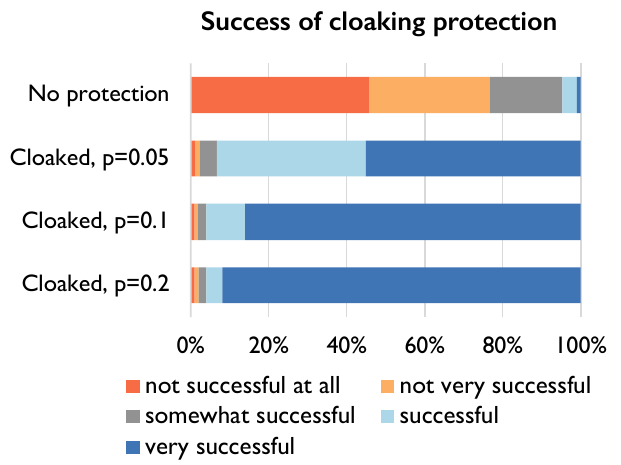}
  \vspace{-0.23in}
  \caption{\system{}'s cloaking protection success increases as cloak perturbation budget increases. The top row of the figure shows baseline performance with the mimic trains on uncloaked images (p=0). }
  \label{fig:budget-increase}
  \end{minipage}
  \hfill
    \centering
    \begin{minipage}{0.32\textwidth}
  \vspace{0.23in}
  \centering
    \resizebox{1\textwidth}{!}{
    \begin{tabular}{lcc}
    \toprule
    \multicolumn{1}{c}{\textbf{\begin{tabular}[c]{@{}c@{}}Perturbation\\ budget\end{tabular}}} & \textbf{\begin{tabular}[c]{@{}c@{}}Artist-rated \\ PSR\end{tabular}} & \textbf{\begin{tabular}[c]{@{}c@{}}CLIP-based \\ genre shift\end{tabular}} \\ \midrule
    0 (no cloak) & $4.6 \pm 1.4\%$ & $2.4 \pm 0.8\%$ \\
    0.05 & $93.3 \pm 0.6\%$ & $96.0 \pm 0.3\%$ \\
    0.1 & $95.9 \pm 0.4\%$& $98.2 \pm 0.1\%$ \\
    0.2 & $96.1 \pm 0.3\%$ & $98.5 \pm 0.1\%$ \\ \bottomrule
    \end{tabular}
    }
    \vspace{0.13in}
    \captionof{table}{Performance of our system (artist-rated protection success rate and CLIP-based genre shift rate) increases as the perturbation budget increases. (SD model, averaged over all victim artists). }
    \label{tab:budget-increase-sd}
  \end{minipage}
  \hfill
\centering
  \begin{minipage}{0.32\textwidth}
  \centering
  \includegraphics[width=1\columnwidth]{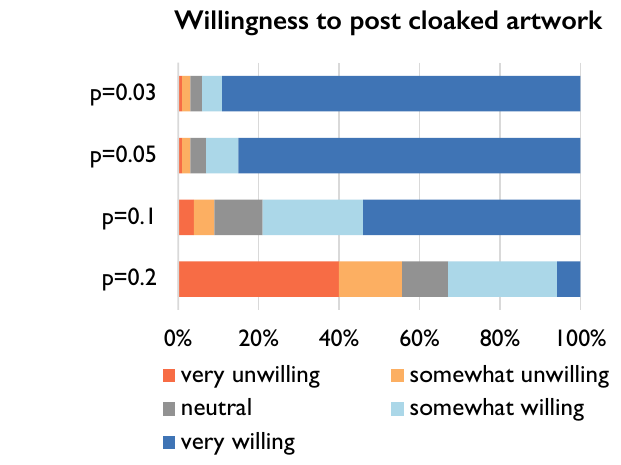}
  \vspace{-0.23in}
  \caption{Artists' willingness to post cloaked artwork in place of the original decreases as perturbation budget of the cloaks increases. }
  \label{fig:artist-accept} 
  \end{minipage}
    \hfill
\end{figure*}

\begin{figure}[t]
  \centering
  \includegraphics[width=0.90\columnwidth]{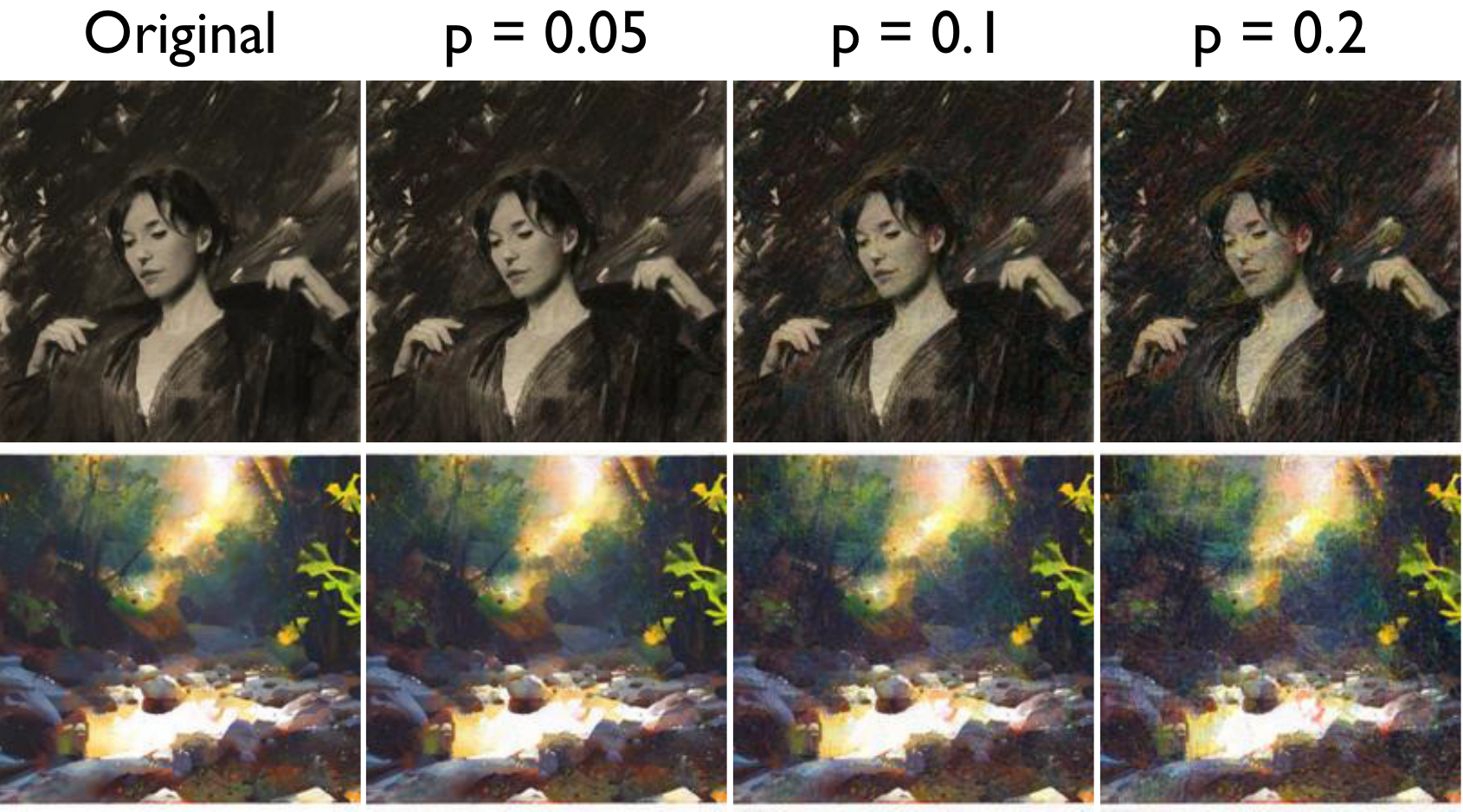}
  \vspace{-0.08in}
  \caption{Original artwork and cloaked artwork computed using three different cloak perturbation budgets. }
  \label{fig:before-after}
\end{figure}

\secspace
\subsection{\system{}'s Protection Performance}
\label{sec:cloaking-results}

\para{Style mimicry success when \system{} is not used. } Mimicry attacks are
very successful when the mimic has access to a victim's original (unmodified)
artwork. Examples of mimicked artwork can be found in
Figure~\ref{fig:core-res}. The leftmost two columns of Figure~\ref{fig:core-res} show a
victim artist's original artwork, while the third column depicts mimicked
artwork generated by a style-specific model trained on victim's original
artwork when \system{} is not used. In our user study, over $>95\%$ of
respondents rated the attack as successful. Table~\ref{tab:psr-core-table},
row 1, gives the artist-rated and CLIP-based genre shift for mimicry attacks on
unprotected art. 

SD models produce stronger mimicry attacks than \dalleM{} models, according
to our user study (see Table~\ref{tab:psr-core-table}). This is unsurprising,
as \dalleM{} models generally produce lower-quality generated
images. CLIP-based genre shift does not reflect this phenomenon, as this metric does
not assess image quality.  

\para{\system{}'s success at preventing style mimicry. } \system{} makes
mimicry attacks markedly less successful, as shown in
Figure~\ref{fig:core-res}. Columns 5 and 6 (from left) show mimicked artwork
when the style-specific models are trained on artwork protected by
\system{}. For reference, column 4 shows an example style-transferred artwork
$\Omega(x, T)$ used to compute \system{} cloaks for the protected art
pieces. Overall, \system{} achieves $> 93.3\%$ artist-rated PSR and
$> 96.0\%$ CLIP-based genre shift (see Table~\ref{tab:psr-core-table}). \system{}'s
protection performance is slightly higher for current artists than for
historical artists. This is likely because the historical artists' images are
present in the training datasets of our generic models (SD, \dalleM),
highlighting the additional challenge of protecting well-known artists whose
style was already learned by the generic models.

\para{How large of perturbations will artists tolerate?} Increasing the
\system{} perturbation budget enhances protection performance. We observe
that both artist-rated and CLIP-based genre shift increase with perturbation budget
(see Figure~\ref{fig:budget-increase}, Table~\ref{tab:budget-increase-sd},
and Figure~\ref{fig:budget2results}). Given this tradeoff between protection
success and \system{} protection visibility on original artwork, we evaluate
how perturbation size impacts artists' willingness to use \system{}. 

We find that artists are willing to add fairly large \system{} perturbations
to their artwork in exchange for protection against mimicry. To measure this,
we show $3$ randomly chosen pairs of original/cloaked artwork to each of the
1,156 artists in our first study. For each art pair, we ask the artist
whether they would be willing to post the cloaked artwork (instead of the
original, unmodified version) on their personal website. More than $92\%$ of
artists select ``willing'' or ``very willing'' when $p=0.05$. This number
only slightly increases to $94.3\%$  when $p=0.03$.
Figure~\ref{fig:artist-accept} details artists' preferences as perturbation
budget increases. (see Figure~\ref{fig:before-after} for examples of cloaked
artwork with increasing $p$). Based on these results, we use perturbation
budget $p = 0.05$ for all our experiments, since most artists are willing to
tolerate this perturbation size.  

Surprisingly, over $32.8\%$ artists are willing to use cloaks with $p=0.2$,
which is clearly visible to human eye (see Figure~\ref{fig:before-after}). While we
are surprised by this high perturbation tolerance, in our follow-up free
response artists noted that they would be willing to tolerate large
perturbations because of the devastating consequence if their styles are
stolen. One participant stated that ``I am willing to sacrifice a bit image
quality for protection.'' Many artists ($>80\%$) also noted that they have
already used traditional, more visually disruptive techniques to protect
their artwork online when posting online, \ie adding watermark or reducing
image resolution. One participant stated that ``I already use low to medium
resolution images only for online posting, thus this would not impact my
quality control too much.'' 

\begin{figure*}[t]
  \centering
  \includegraphics[width=0.95\linewidth]{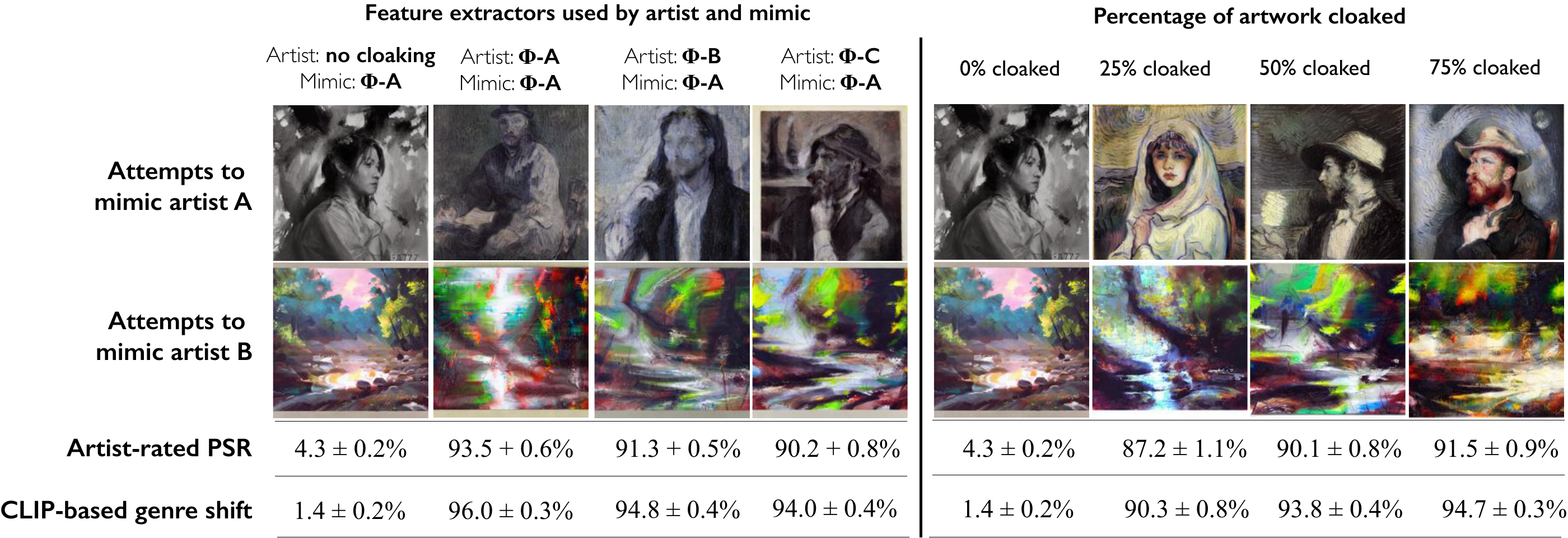}
  \caption{\system{} remains successful under two challenging
    scenarios. Left: when artist and mimic use different feature
    extractors. Right: when artists can only cloak a portion of their artwork
    in mimic's dataset. Bottom of the figure shows artist-rated PSR and
    CLIP-based genre shift for the corresponding setting. } 
  \label{fig:core-robust}
\end{figure*}

\secspace
\subsection{\system{}'s Protection Robustness}
\label{sec:robust-eval}

\secspace

Next, we test \system{}'s efficacy in more challenging scenarios. First, we
measure performance when the mimic uses a different feature extractor for
mimicry than the one used by the artist to generate the cloak. Second, we
measure what happens when the mimic has uncloaked artwork samples from the
victim.  Due to the poor mimicry performance of \dalleM, we focus our
evaluation using SD as the generic model.

\para{Artist/mimic use different feature extractors. } In the real world, it
is possible that the mimic will use a different model (and thus a different image
feature extractor) for style mimicry than the one used by the victim artist
to cloak their artwork. While the feature extractors may still be similar
because of the well-known transferability property between large
models~\cite{demontis2019adversarial,transfer,suciu2018does,transfer2014,shan2022post},
their differences could reduce the efficacy of cloaking. We test this
scenario using three feature extractors\textemdash $\Phi$-A, $\Phi$-B, and
$\Phi$-C. $\Phi$-A and $\Phi$-B have different model architectures
(autoencoder-KL~\cite{rombach2022high} vs. VQ-VAE~\cite{ramesh2021zero}) but
are both trained on the ImageNet dataset~\cite{deng2009imagenet}. $\Phi$-A
and $\Phi$-C have different model architectures (autoencoder-KL vs VQ-VAE)
and training datasets (ImageNet vs. CelebA~\cite{liu2018large}).

In our experiments, the victim artist uses one feature extractor (either
$\Phi$-B or $\Phi$-C) to optimize cloaked artwork, and the mimic trains their
style-specific models with SD models using $\Phi$-A. Despite the difference
in victim/mimic extractors, \system{}'s protection remains highly successful
(left half of Figure~\ref{fig:core-robust})\textemdash the style of mimicked
artwork remains distinct from artist's true style. Artist-rated and
CLIP-based genre shift measurements confirm this observation. Artist-rated PSR is
$> 90.2\%$, while CLIP-based genre shift is $> 94.0\%$. The PSR is slightly higher
when the two feature extractors only differ in architectures ($\Phi$-B to
$\Phi$-A) than when they differ in both architecture and training data
($\Phi$-C to $\Phi$-A).

\para{Mimic has access to uncloaked artwork. } Another challenging scenario
is when the mimic gains access to some \textit{uncloaked} artwork from victim
artists. This is a realistic scenario for many prominent artists with a large
online presence. As expected, \system{}'s protection performance decreases
when the mimic has access to more uncloaked artwork (right side of
Figure~\ref{fig:core-robust}). As the ratio of uncloaked/cloaked art in the
mimic's dataset increases, the mimicked artwork becomes more similar to
artist's original style. Yet, \system{} is still reasonably effective
($87.2\%$ artist-rated PSR) even when artists can only cloak $25\%$ of their
artwork. This validates our hypothesis in \S\ref{sec:cloak-effect} that
cloaking will have a noticeable effect as long as the mimic has some cloaked
training data.

A mimic with access to a large amount of uncloaked artwork is still an issue
for \system{}. Fortunately, in our user study, we found that 1) many artists
constantly create and share new artwork online, which can be cloaked to
offset the percentage of uncloaked artwork, and 2) many artists change their
artistic style over time. In our user study, we asked artists to estimate the
number of unique art pieces they currently have online ($M$) and the
estimated number of art pieces they anticipate uploading each subsequent year
($Y$). Among artists with an existing online presence, over $40\%$ have
$Y / M > 25\%$, meaning that one year from now, $> 20\%$ of their total
online artwork would be cloaked (if they start using \system{}
immediately). More than $81\%$ of artists also stated that their art style
has changed over their career, and half of these said that theft of their
old, outdated styles is less concerning.

\begin{table}[t]
  \centering
  \resizebox{0.49\textwidth}{!}{
  \centering
  \begin{tabular}{ccccc}
    \hline
    \multirow{2}{*}{\textbf{\begin{tabular}[c]{@{}c@{}}Artist \\ dataset\end{tabular}}} & \multicolumn{2}{c}{\textbf{w/o \system}} & \multicolumn{2}{c}{\textbf{w/ \system{} (p=0.05)}} \\ \cline{2-5} 
     & \begin{tabular}[c]{@{}c@{}}Artist-rated \\ PSR\end{tabular} & \begin{tabular}[c]{@{}c@{}}CLIP-based \\ genre shift\end{tabular} & \begin{tabular}[c]{@{}c@{}}Artist-rated \\ PSR\end{tabular} & \begin{tabular}[c]{@{}c@{}}CLIP-based \\ genre shift\end{tabular} \\ \hline

    Current & $6.2 \pm 0.5\%$ & $3.8 \pm 0.3\%$ & $92.5 \pm 0.5\%$ & $94.2 + 0.3\%$ \\
    Historical & $7.2 \pm 0.6\%$ & $3.3 \pm 0.4\%$ & $92.1 + 0.3\%$ & $93.9 + 0.4\%$ \\ 
    \hline
    \end{tabular}
  }
  \vspace{-0.1in}
  \caption{Performance of \system{} against real-world mimicry service
    (scenario.gg). Mimicry service achieves high mimicry success when no
    protection is used. When \system{} is used, the mimicry service has low
    performance. }
  \label{tab:real-world}
\end{table}

\secspace
\subsection{Real-World Performance}

Next, we test \system{} against a real-world style mimicry-as-a-service
system, \texttt{scenario.gg}~\cite{aigame}. Scenario.gg is a web service that
allows users to upload a set of images in a specific style. The
service then trains a model to mimic the style and returns an API endpoint
that allows the user to generate mimicked images in the trained style. The
type of model or mimicry method used by the service is unknown.

\system{} remains effective against \texttt{scenario.gg}. We ask
\texttt{scenario.gg} to mimic the style from a set of cloaked or uncloaked
artwork from $4$ current artists and $19$ historical
artists. Table~\ref{tab:real-world} shows that when no protection is used,
\texttt{scenario.gg} can successfully mimic the victim style (< 7.2\%
protection success). The mimicry success of \texttt{scenario.gg} is lower
than our mimicry technique, likely because \texttt{scenario.gg} trains the
model for fewer iterations due to computational constraints. When we use
\system{} to cloak the artwork and upload the cloaked artwork,
\texttt{scenario.gg} fails to mimic the victim style ($> 92.1\%$ artist-rated
PSR and $> 93.9\%$ CLIP-based genre shift rate) as shown in Table~\ref{tab:real-world}.

\begin{figure}[t]
  \centering
  \includegraphics[width=1\columnwidth]{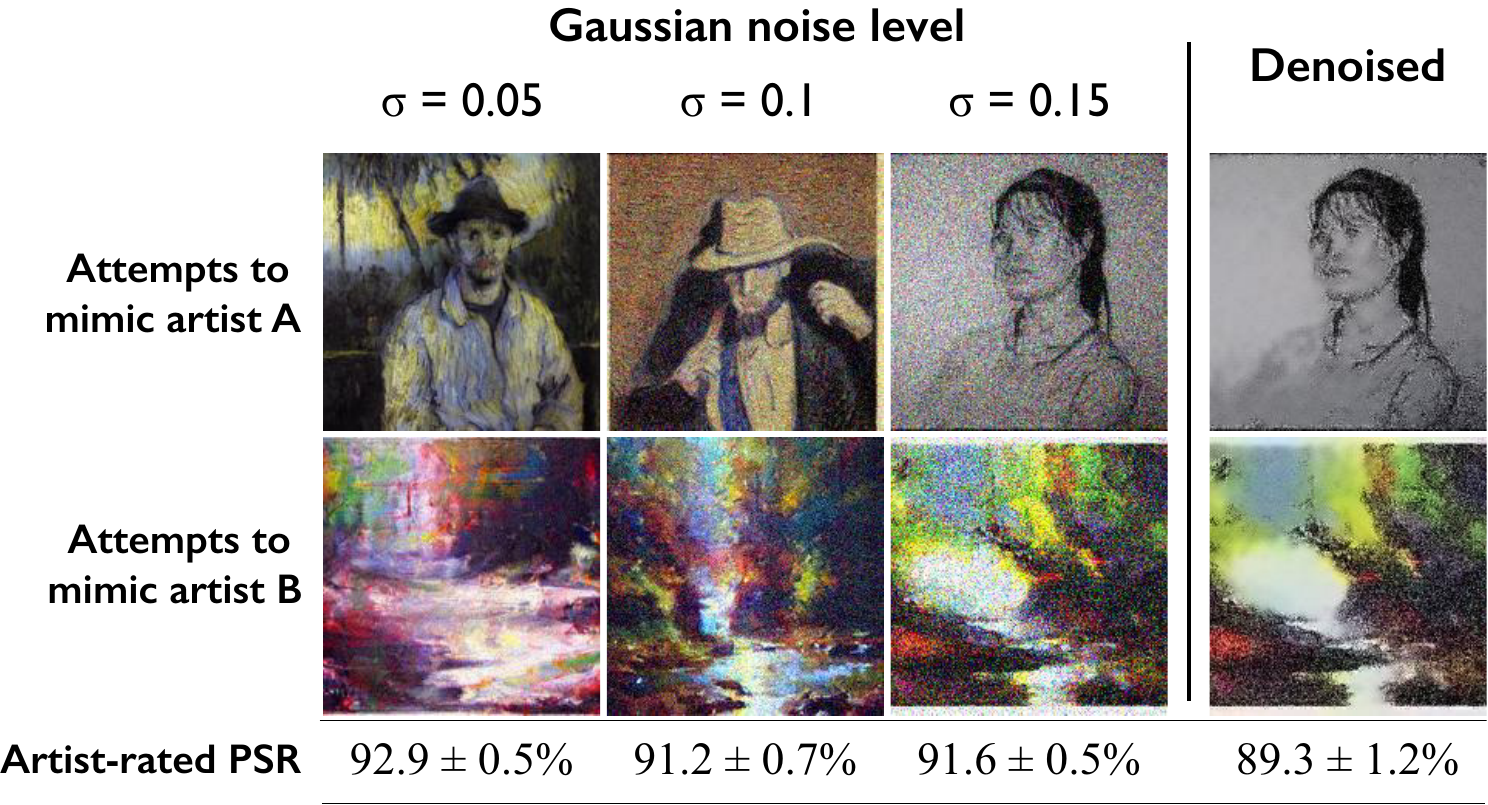}
    \vspace{-0.25in}
  \caption{\system{}'s protection performance remains high as mimic adds an
    increasing amount of Gaussian noise to the cloaked artwork. Even when the
    mimic adds denoising (last column), \system{}'s protection persists. } 
  \label{fig:noise_countermeasure}
\end{figure}

\begin{figure}[t]
  \centering
  \includegraphics[width=1\columnwidth]{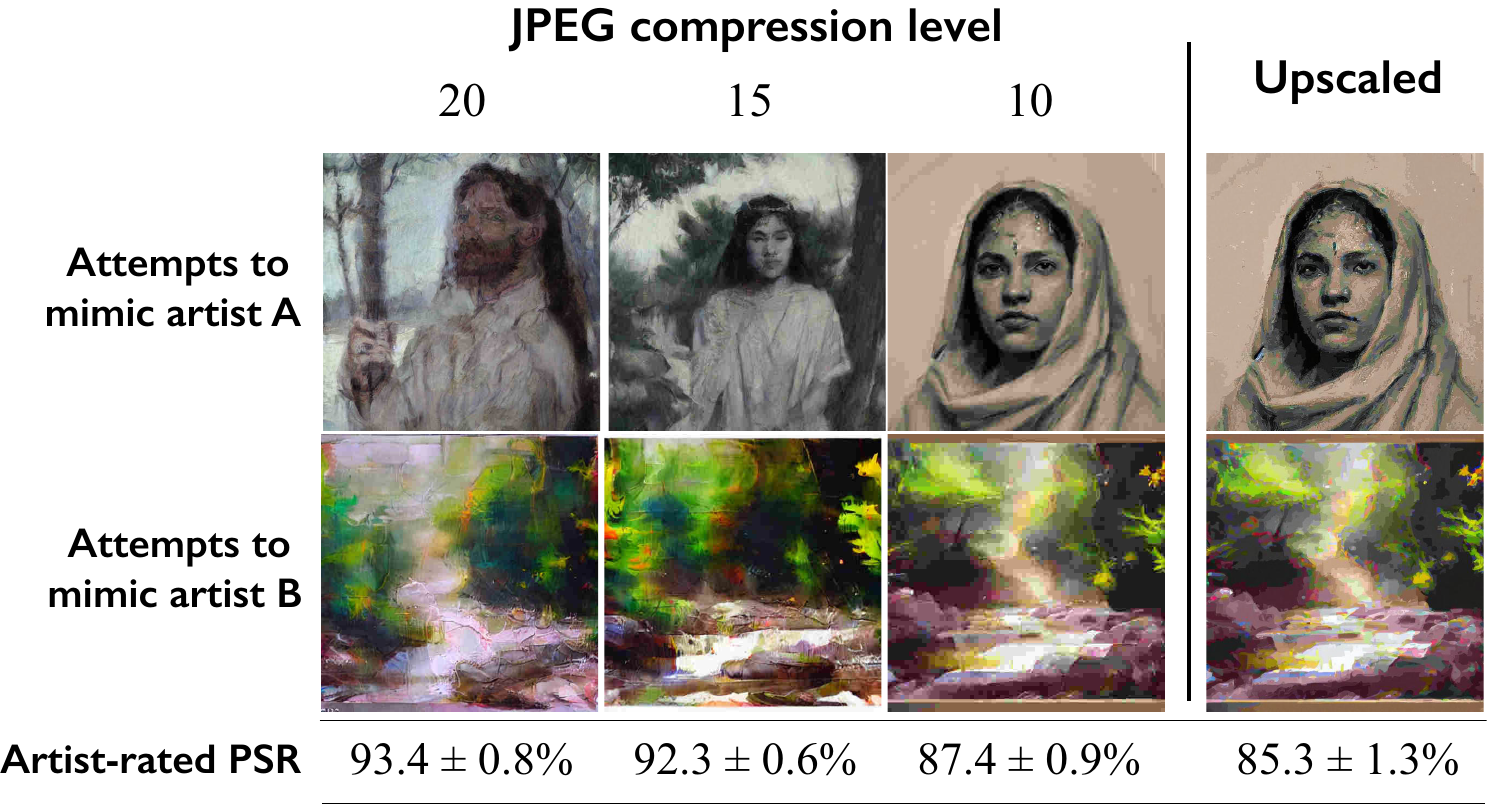}
  \vspace{-0.25in}
  \caption{\system{}'s protection performance remains high as mimic adds JPEG compression to the cloaked artwork. Even when the mimic also upscales the mimicked images (last column), \system{}'s protection persists. }
  \label{fig:jpeg_countermeasure}
\end{figure}

\secspace
\section{Countermeasures} 
\label{sec:counter}

We consider potential countermeasures a mimic could employ to reduce the
effectiveness of \system. We consider the strongest adversarial setting, in
which the mimic has white-box access to our protection system, \ie access to
the feature extractor used and protection algorithm. In our experiments, we
assume the mimic uses the SD model as the generic model and test the efficacy
of each countermeasure on the $13$ victim artists from
\S\ref{sec:metrics}. Here, we focus on artist-rated PSR metric, because many
countermeasures trade off image quality for mimicry efficacy, and
CLIP-based metric does not consider image quality.

\para{Image transformation. } A popular approach to mitigate the impact of
small image perturbations, like those introduced by \system{}, is to
transform training images before using them for model
training~\cite{carlini2017adversarial,feinman2017detecting}. In our setting,
the mimic could augment the cloaked artwork before fine-tuning their model on
them to potentially reduce cloak efficacy. We first test \system{}'s
resistance to two popular image transformations, adding Gaussian noise and
image compression. We also consider a stronger version of this countermeasure
that then tries to correct the image quality degradation introduced by the
transformations.

Transforming cloaked artwork does not defeat \system{}'s
protection. Figure~\ref{fig:noise_countermeasure} shows that as the magnitude
of Gaussian noise ($\sigma$) increases, the quality of mimicked artwork
decreases as fast as or faster than cloak effectiveness. This is because
models trained on noisy images learn to generate noisy images. We observe a
similar outcome when mimic uses JPEG compression
(Figure~\ref{fig:jpeg_countermeasure}), where image resolution and quality
degrade due to heavy compression. Artists-rated PSR decreases slightly but
remains above $>87.4\%$ across both types of data transformations. Artists
consider \system{}'s protection to be successful when mimicked artwork is of
poor quality.  

The mimic can take this countermeasure one step further by \textit{reversing}
the quality degradation introduced by the noising/compression
process. Specifically, a mimic can run image denoising or image upscaling
tools on the mimicked artwork (\eg ones shown in
Figure~\ref{fig:noise_countermeasure} and \ref{fig:jpeg_countermeasure}) to
increase their quality. We found this approach improves generated image
quality but still does not allow for successful mimicry. For denoising, we
ran a state-of-the-art CNN-based image denoiser~\cite{zhang2017beyond} that
is specifically trained to remove ``additive Gaussian noise'' (the same type
of noise added to cloaked artwork). The last column of
Figure~\ref{fig:noise_countermeasure} shows the denoised image (using the
noisy mimicked image when $\sigma=0.2$ as the input). While the process
removes significant amounts of noise, the denoised artwork still has many
artifacts, especially around complex areas of the artwork (\eg human
face). We observe similar results for image upscaling, where we use a
diffusion-based image upscaler~\cite{stable2-1} to improve the quality of
compressed images (Figure~\ref{fig:jpeg_countermeasure}). Overall, our
artist-rated protection success rate remains $> 85.3\%$ against this improved
countermeasure.  

\begin{figure}[t]
  \centering
  \includegraphics[width=1\columnwidth]{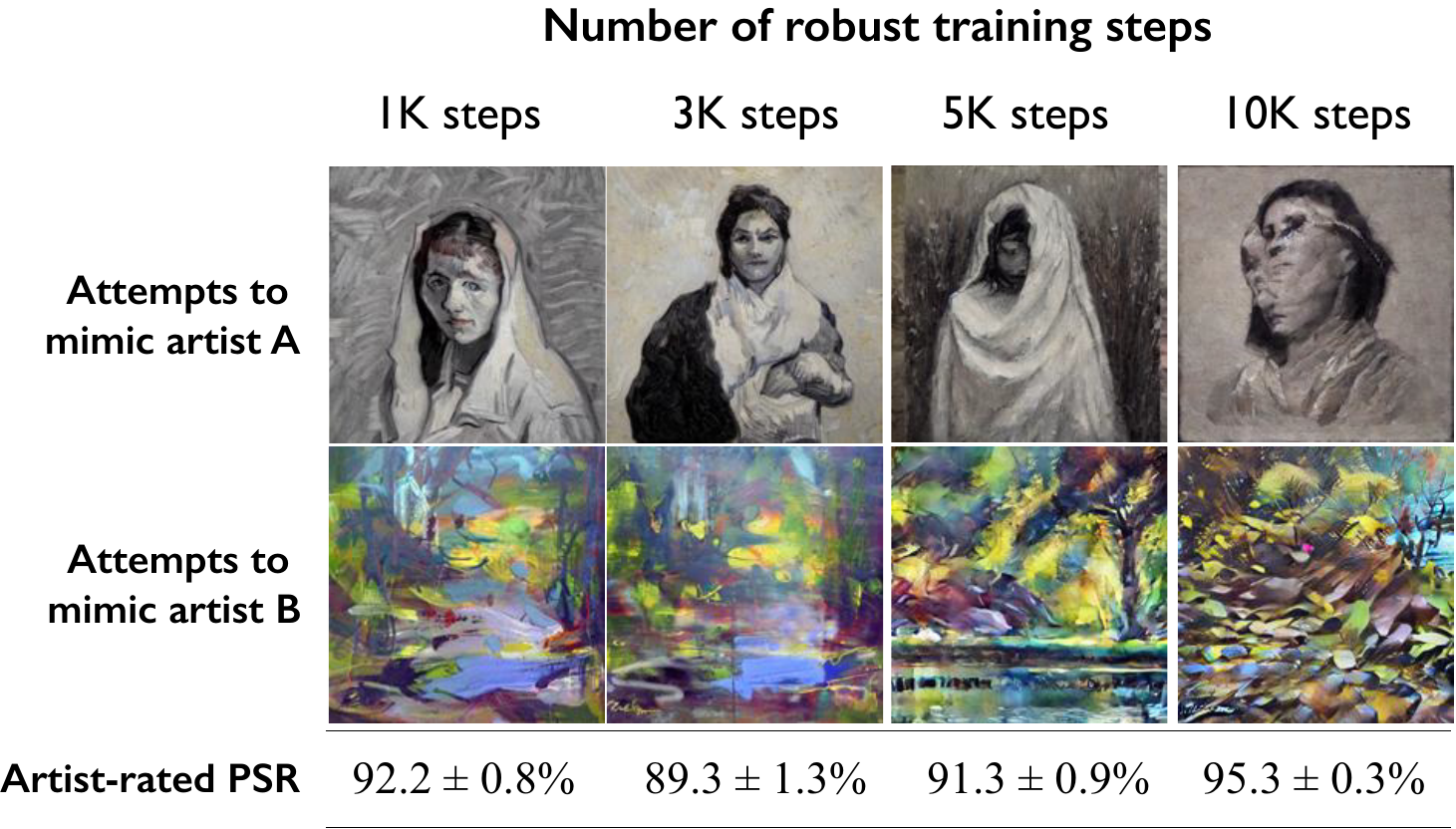}
  \caption{\system{}'s protection performance remains high against robust training countermeasure proposed by Radiya \etal. The protection performance first decreases then increases as mimic robustly trains the model with an increasing number of steps. }
  \label{fig:florian}
\end{figure}

\para{Radiya \etal~\cite{radiya2021data} robust training.} Radiya
\etal~\cite{radiya2021data} design a robust training method to defeat
cloaking tools like Fawkes~\cite{shan2020fawkes} and
Lowkey~\cite{cherepanova2021lowkey} in the face recognition setting. At a
high level, this method augments the attacker's training dataset with some
cloaked images generated by the cloaking tool and the \textit{correct} output
labels. Training on such data makes the model more robust against cloak
perturbations on unseen cloaked images at inference time, and thus, can
potentially circumvent the protection.


We test if this robust training approach can defeat \system{}. We assume the
mimic first robustly trains the feature extractors in their generic models
using cloaked artwork generated by \system{}, and then trains the generator
model to generate images from the robust feature space. Finally, the mimic
uses the robust generic model for style mimicry as in
\S\ref{sec:eval-cloak}. We discuss the detailed robust training setup in
Appendix~\ref{app:counter}.  

\system{} performance remains high, even if the mimic robustly trains the
generic model for many iterations before using it for style mimicry (see
Figure~\ref{fig:florian}). As the model becomes more robust, the mimicked
artwork is less impacted by cloaking (less influenced of the target
style). However, robust training greatly degrades mimicked image quality,
preventing successful mimicry. Overall, the artist-rated PSR remains
$>~88.7\%$. To mitigate robust training's impact on image quality, we explore
an alternative robust training method, where we robustly train a new feature
extractor designed to remove cloak's impact while operating in the
original feature space (thus no need to change the image
generator). We found this robust training approach is also ineffective
(details in \S\ref{app:counter}).

As discussed in \S\ref{sec:cloak-effect}, \system{} remains reasonably
effective against Radiya \etal because 1) the continuous output space of the
generative model, and 2) high quality requirement of art generation. Robust
training reduces cloaking's effectiveness but cannot completely remove its
impact. In the classification case (facial recognition), this reduced
effectiveness only manifests in small changes in classification confidence
(compared to no cloaking) and often does not change the discrete classification
outcome. However, in the context of generator models, the continuous output space means
that even less-effective cloaks still directly affect the mimicked
artwork. Combined with the high quality requirement, the reduced protection
effect is enough to disrupt style mimicry, as shown in
Figure~\ref{fig:florian}. Additional robust training simply degrades
generation quality, rather than reducing cloaking efficacy.

\revise{
  
\para{Outlier Detection. } Another countermeasure could involve
leveraging outlier detection to identify and remove protected images~\cite{wang2021understanding,shan2020gotta,wang2019neural}. We
test \system{}'s robustness to a state-of-the-art outlier detection method that leverages
contrastive training~\cite{wang2021understanding}. Contrastively trained
models project data into a well-separated feature space, which the mimic
could leverage. 

We assume the mimic has a ground truth set (20) of original artworks from a
given artist. The mimic first projects these art pieces into the feature space
of a model trained with contrastive loss on ImageNet
dataset~\cite{wang2021understanding}. The mimic then trains a one-class SVM
outlier detector~\cite{li2003improving} using these ground truth
features. Now, given a new artwork from the same artists, the mimic detects
whether the artwork is an outlier using the detector. Detection results on
$4$ current artists (\S\ref{sec:eval-cloak}) show that outlier detection has
limited effectiveness against \system{} ($< 65\%$ precision and $< 53\%$
recall at detecting \system{} protected images).  
}

\vspace{-0.1in}
\section{Limitations and Releasing Glaze}
We conclude with a discussion of the limitations of the current system,
then describe our experiences during and after the \system{} release.

\para{Limitations.}  First, protection from \system{} relies on artists cloaking
a portion of their art in the mimic model's training dataset. This is
challenging for established artists because 1) their styles have matured over
the years and are more stable, and 2) many of their art pieces have already
been downloaded from art repositories like ArtStation and DeviantArt. These
artists' styles can be mimicked using only older artworks collected before
the release of \system.  While artists can prevent mimics from training on
newer artwork, they need to rely on opt-out and removal options at art
repositories to stop style mimicry.

Second, a system like \system{} that protects artists faces an inherent
challenge of being {\em future-proof}. Any technique we use to cloak artworks
today might be overcome by a future countermeasure, possibly rendering
previously protected art vulnerable. While we are under no illusion that
\system{} will remain future-proof in the long run, we believe it is an important and
necessary first step towards artist-centric protection tools to
resist invasive AI mimicry. We hope that \system{} and followup projects 
will provide some protection to artists while longer term (legal, regulatory)
efforts take hold.

\revise{ \para{Releasing \system{} and managing expectations. } We released
  \system{} as a free application on Mac and Windows in March 2023. We have
  repeatedly communicated \system{}'s limitations to users, 
  both on our website and in communications to artists via our
  download page, on Twitter, in emails to artists, etc. In these
  communications, we clearly state that \system{} is not a permanent solution
  against AI mimicry and could potentially be defeated by future attacks.

  As of June 2023, \system{} has been downloaded $>740K$ times by artists
  around the world. Reception on social media and emails to our lab have been
  extremely enthusiastic and positive. Artists have helped design Glaze's user interface, made
  how-to videos on YouTube, and managed ad campaigns on Instagram to spur
  adoption in the community. Based on numerous requests on social media and
  via emails, we plan to test and deploy a web service in Summer 2023
  to expand \system{} access to artists who lack compute and GPUs.

  One excellent outcome from the \system{} release has been the technical
  discussions it has spurred with a variety of stakeholders. We began and are
  continuing collaborative efforts to advocate for artists rights, with
  art-centric social networks, advocate groups in the US (CAA) and the EU
  (EGAIR), government representatives, and companies who want to protect the
  IP of their images/characters.}

\para{Real-world countermeasures.} Finally, we want to describe our
experiences deploying \system{} in an adversarial setting. In the 3 months
since initial release, multiple groups have sought to attack or bypass \system{}
protection. While several attempts had minimal impact, we describe the two
most serious attempts here and evaluate their effectiveness.

\begin{figure}[t]
  \centering
  \includegraphics[width=0.90\columnwidth]{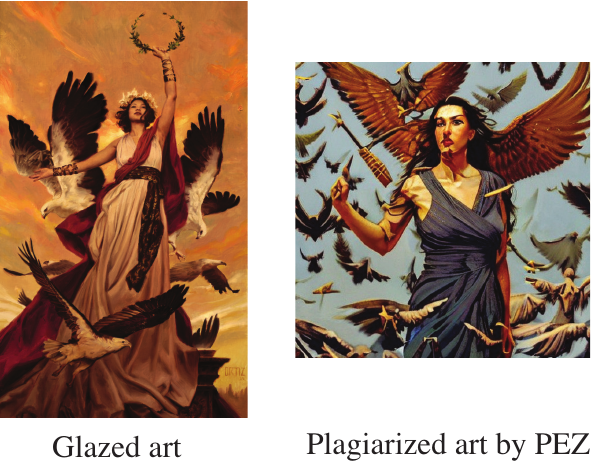}
  \vspace{-0.1in}
  \caption{Glazed image and generated image from PEZ mimicry method. The
    original image is {\em Musa Victoriosa}, a new painting created by Karla
    Ortiz to be the first artwork to be released publicly under \system{} protection.}
  \label{fig:marx-attack}
\end{figure}

The first attack~\cite{marx-attack} leverages a newer style mimicry
method~\cite{wen2023hard}, reverse engineering with PEZ. PEZ is able to
perform high-quality style mimicry using \textit{a single original image}
from the original artist. Initial tests showed \system{} is robust
against PEZ style mimicry (Figure~\ref{fig:marx-attack}). \system{}
remains effective likely because \system{} directly modifies the
feature representation of the art, and is thus effective against stronger mimicry
attempts.

\begin{figure}[t]
  \centering
  \includegraphics[width=0.90\columnwidth]{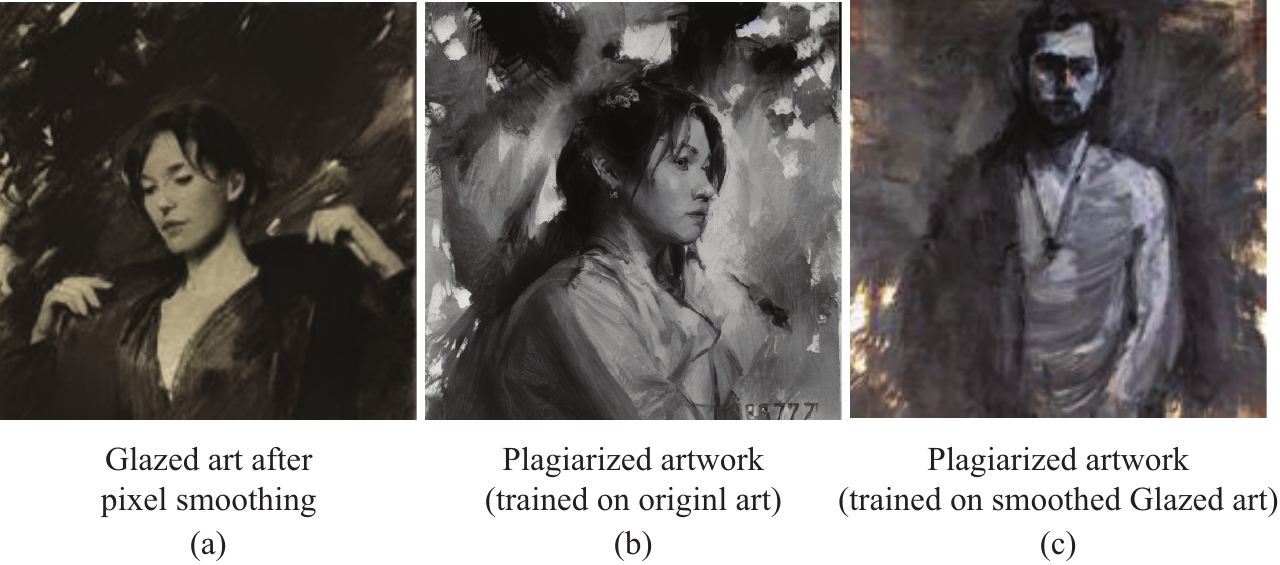}
  \vspace{-0.1in}
  \caption{ (a) Smoothed artwork by applying pixel smoother on Glazed
    artwork, (b) plagiarized artwork generated by training on original
    (unprotected) artwork, and (c) plagiarized artwork generated by training
    on Glazed artwork that was later pixel-smoothed.} 
  \label{fig:smooth-attack}
\end{figure}

A second category of attacks tries to perform pixel-level image smoothing to remove
cloaks added by \system{}~\cite{smooth-attack}. This applies bilateral
filters on Glazed images repeatedly, seeking to remove all added perturbations. We
evaluate this attack on Glazed artwork in \S\ref{sec:eval-cloak} and
fine-tuning a model on the smoothed images. Figure~\ref{fig:smooth-attack}
shows \system{} remains effective against pixel smoothing. This result is
consistent with prior work showing that image smoothing cannot prevent
adversarial perturbations~\cite{zhang2020smooth}.

Finally, while we have not yet observed any successful attacks against \system{},
we are continuously exploring design improvements to further enhance
robustness against potential future countermeasures.

\section*{Acknowledgements} We thank our anonymous reviewers and shepherd for
their insightful feedback. We also thank Karla Ortiz, Lyndsey Gallant, Nathan
Fowkes, Kim Van Deun, Jon Lam, Eveline Fr\"ohlich, Edit Ballai, Kat Loveland and many
other artists, without whom this project would not be possible. This work is
supported in part by NSF grants CNS-2241303, CNS-1949650, and the DARPA GARD
program.  Opinions, findings, and conclusions or recommendations expressed in
this material are those of the authors and do not necessarily reflect the
views of any funding agencies.

{
 \footnotesize
 \bibliographystyle{acm}
 \bibliography{glaze}
}

\balance

\appendix
\section{Appendix}
\label{sec:appendix}

\begin{figure*}[t]
\centering
  \begin{minipage}{0.31\textwidth}
  \centering
  \includegraphics[width=1\columnwidth]{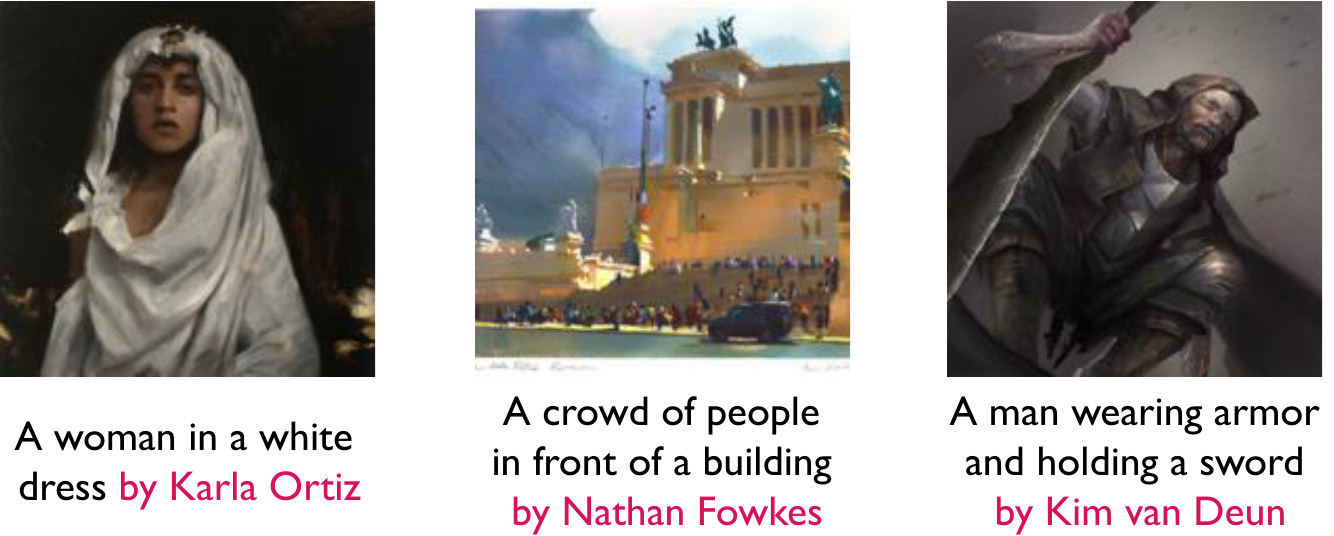}
  \caption{Example data used for fine-tuning, including artwork from different
    artists and their text captions.} 
  \label{fig:data-examples}
  \end{minipage}
\hfill
  \begin{minipage}{0.31\textwidth}
  \centering
  \includegraphics[width=1\columnwidth]{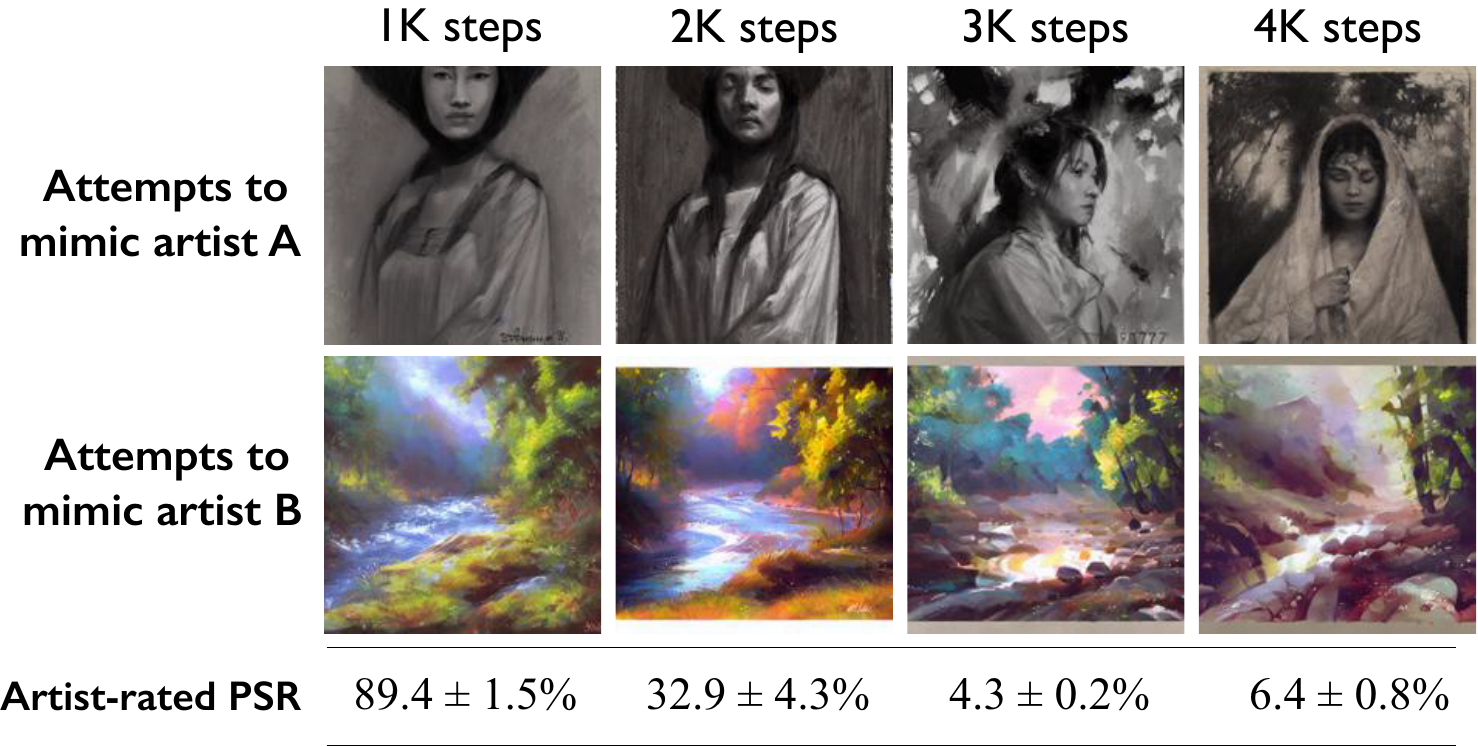}
  \caption{The success of style mimicry when the mimic fine-tunes the model for an increasing number of iterations. }
  \label{fig:success-iter}
  \end{minipage}
\hfill
  \begin{minipage}{0.31\textwidth}
  \centering
  \includegraphics[width=1\columnwidth]{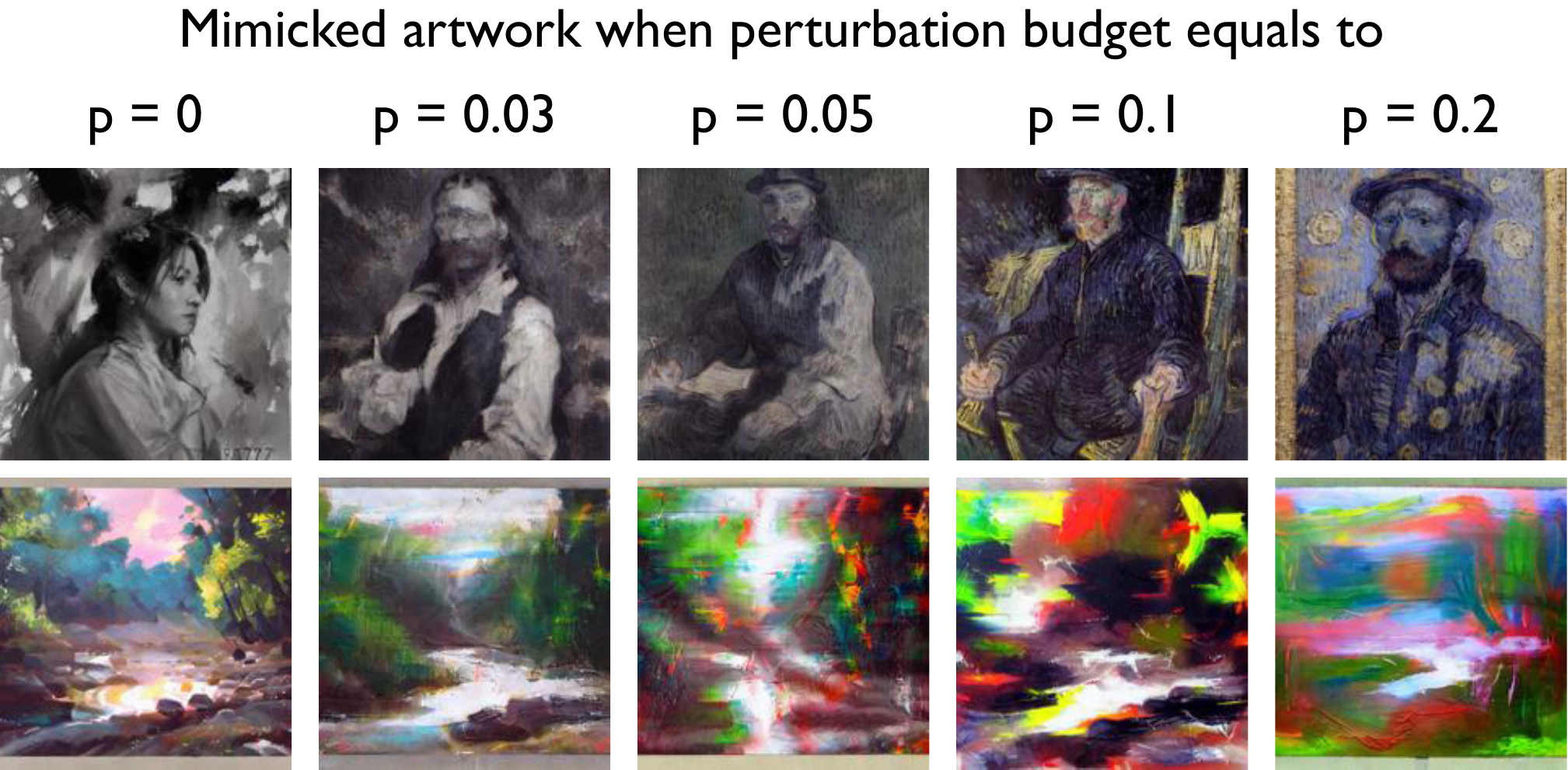}
  \caption{Mimicked artwork when artist uses an increasingly high
    perturbation budget to protect their original art.} 
  \label{fig:budget2results}
  \end{minipage}
  \hfill

\end{figure*}

\begin{figure*}[t]
  \begin{minipage}{0.44\textwidth}
  \centering
  \includegraphics[width=1\columnwidth]{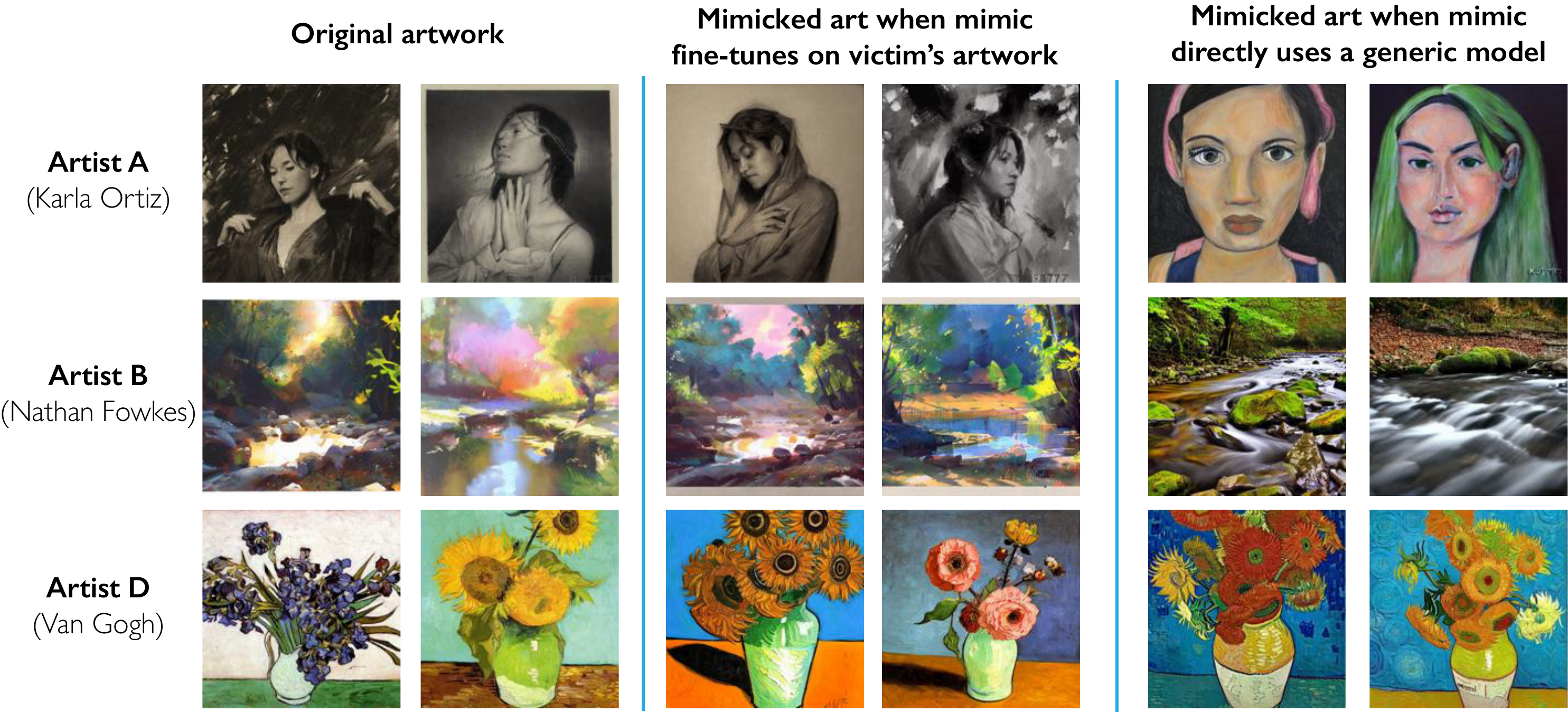}
  \vspace{-0.23in}
  \caption{Comparing performance of art mimicry directly using a generic
    model to that of mimicry on a model that has been fine-tuned on the
    victim's art pieces. \textbf{Column 1-2}: artists' original
    artwork; \textbf{column 3-4}: plagiarized artwork generated from a
    style-specific model fine-tuned on artist's art; \textbf{column
      5-6}: plagiarized artwork generated from the generic SD model using the
    artist's name as prompt.}
  \label{fig:generic-vs-finetune}
  \end{minipage}
  \hfill
\centering
  \begin{minipage}{0.48\textwidth}
  \centering
  \includegraphics[width=1\columnwidth]{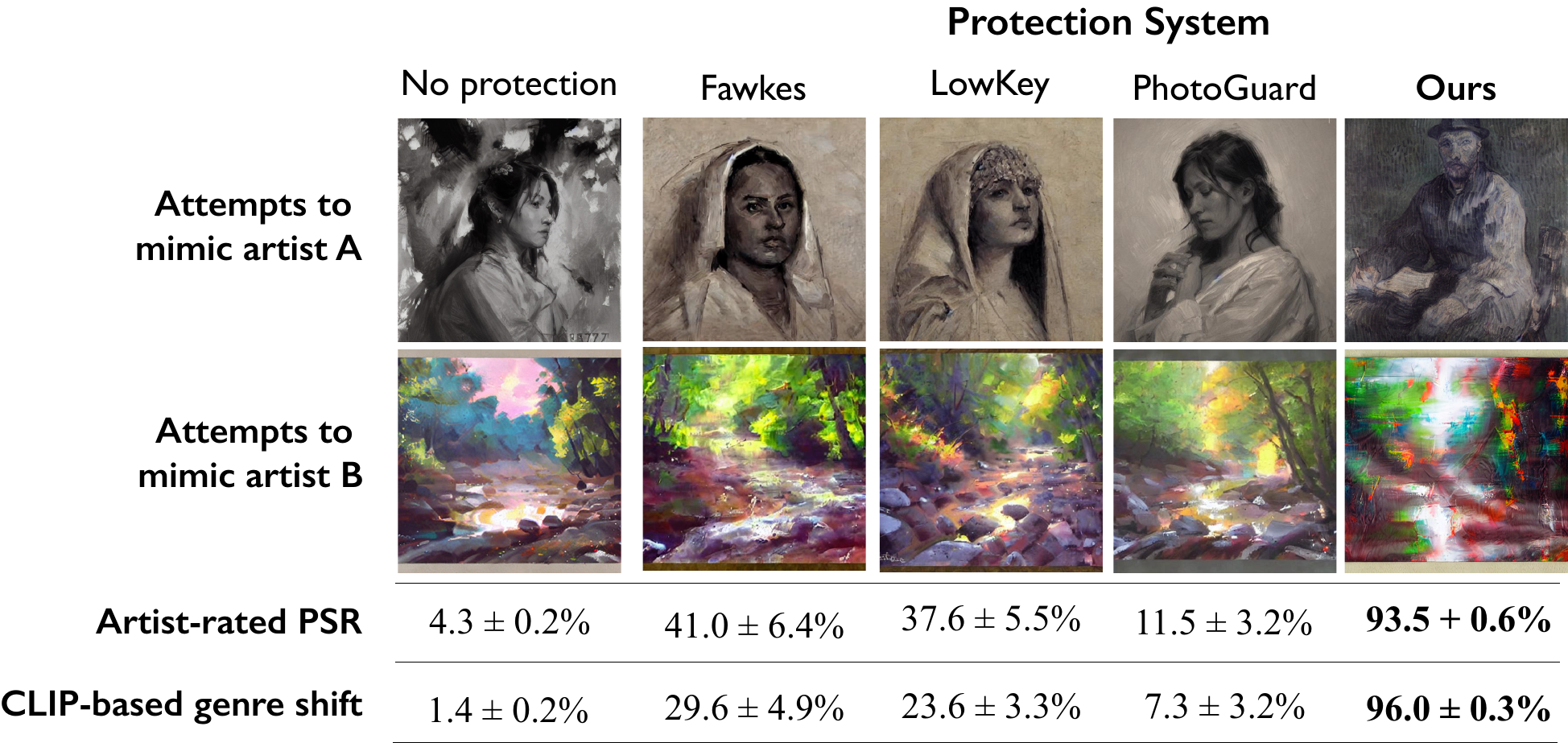}
  \vspace{-0.23in}
  \caption{Comparing protection levels provided by different cloaking
    systems, including adapted versions of Fawkes, Lowkey, and Photoguard for style
    protection. \system{} significantly outperforms these adapted alternatives.} 
  \label{fig:existing-sys}
  \end{minipage}
  \hfill
\end{figure*}

\subsection{Adapting Existing Cloaking Systems}
\label{app:prior-work}

Here, we consider whether prior image cloaking systems can be adapted to
provide protection against art style mimicry. Our results show adapting
existing cloaking systems has limited effectiveness for our goals.

\para{Adapting existing cloaking systems. } Fawkes~\cite{shan2020fawkes}
generates a cloak on user face images by optimizing the feature space
difference between the cloaked image and a target image. The target image is
simply a face image of a different person. We adapt Fawkes to anti-mimicry
protection by switching the feature extractor from facial recognition to the
same one we use for \system. For the target image used, we assume Fawkes
randomly picks an artwork from a different artist. Fawkes uses DSSIM to bound
the input perturbation. For a fair comparison, we change Fawkes perturbation
from DSSIM to LPIPS, ones used by \system.

The general design of Lowkey~\cite{cherepanova2021lowkey} is similar to
Fawkes, except Lowkey does not optimize cloak images towards a target in
feature space but simply optimizes cloaked images to be different from the
original one. We directly apply LowKey for anti-mimicry protection: Lowkey
maximizes the cloaked artwork to have a different feature representation from
the original artwork. 

Photoguard~\cite{madry-defense} works by minimizing the norm of the image
feature vector. It is equivalent to Fawkes when Fawkes selects the zero
feature vector as the target for optimization. For anti-mimicry, we adapted
Photoguard to minimize the norm of feature representation of the cloaked
artwork. 

\para{Performance comparison. } Figure~\ref{fig:existing-sys} show Fawkes,
Lowkey, and Photoguard have limited effectiveness at protection against mimicry.
Out of the three existing systems, Fawkes achieves the best
performance with $41.0\%$ artist-rated protection success rate. While we can
see small artifacts introduced by Fawkes and Lowkey, they are not sufficient
to prevent mimicry. In our tests, we use the same LPIPS perturbation level
and the same feature extractor for optimization for all cloaking systems.  

\subsection{Additional Information on Style Mimicry}
\label{app:mimicry}

\para{Impact of fine-tuning on mimicry success. }
Figure~\ref{fig:generic-vs-finetune} compares the mimicry performance when
a mimic attack fine-tunes on the victim artist's artwork or directly using a
generic model. For artists who are not household names (e.g. iconic artists like Van Gogh),
fine-tuning significantly improves mimicry performance. We generate 
images using text captions containing the artist's name, \eg ``a river by
Nathan Fowkes.''   

\para{Details on training parameters. } For stable diffusion, we follow the
same training parameters as the original paper~\cite{rombach2022high}. We use
$5 \cdot 10^{-6}$ learning rate and batch size of $32$. For a generation, we
follow the default setting using the PNDM sampler and $50$ sampling
steps. For \dalleM, we also follow the same training setup as~\cite{d-mini} with
a learning rate of $2 \cdot 10^{-5}$ and batch size $32$. To generate images,
we use the default setting with a condition scale equal to $10$. 

\para{Impact of selecting random seed. } For diffusion-based models (\eg SD),
artwork generation is controlled by a random seed (random noise input at the
beginning). Different random seeds lead to very different images, and thus
it is common practice for mimics to generate a set of artwork using
different seed and select the best artwork. A relevant question is, can a
mimic use sheer randomness to generate a plagiarized artwork that succeeds
despite Glaze protection.

We investigate the impact of random seed selection on mimicry success in the
presence of Glaze. Given a style-specific model and a given text prompt, the
mimic generates $100$ plagiarized artworks using different random
seeds. Similar to how we calculate CLIP-based genre shift, we then use the
CLIP model to identify any artwork that belongs to the same genre as the
target artist's style. The results show that $4.3\%$ of the time, the mimic is able to find
at least $1$ out of the $100$ plagiarized artwork that passed CLIP
filtering. While the filtered artwork does belong to the same genre as the
artist, we found they tend to have lower image quality. We verify this
observation in our user study, and $> 94.1\%$ human artists rated the
protection remains successful (i.e. these art pieces failed to mimick the art
style). We believe the reason that
some plagiarized artwork still shares the same genre as victim style after
protection, is that text-to-image models today are still imperfect and often
output poor-quality images in rare cases with some random seed.

\subsection{CLIP-based metric}
\label{app:clip}

We test CLIP's performance in classifying artwork into the correct art
genre. We take $27$ historical genres from WikiArt and $13$ digital art
genres~\cite{digital-styles} as the candidate labels. We collect a test
dataset consisting of $1000$ artwork from WikiArt dataset, each containing
the ground truth labels from the Wikiart dataset. Then we collect $100$
artwork for each of the $13$ digital art genres by searching the name of the
genre on ArtStation, one of the largest digital art-sharing platforms. We
evaluate CLIP performance using top-3 accuracy as many art genres are similar
to each other (\eg impressionism vs fauvism). CLIP achieves $96.4\%$ top-3
accuracy on artwork from WikiArt and $94.2\%$ for artwork from ArtStation.

\subsection{Additional Countermeasures} 
\label{app:counter}

\para{Details on robust training. } Here, we give details on the robust
training method we used. We follow prior work~\cite{salehi2021arae} on robust
training of autoencoder models. The mimic first uses \system{} to generate a
large number of cloaked artwork using artwork from WikiArt dataset. Given the
feature extractor $\Phi$ used by mimic's text-to-image model, the mimic trains
$\Phi$ to minimize the following loss function:  

 \begin{eqnarray}
   &\min_{\Phi} ||\Phi(x_{cloaked}) - \Phi(x_{org})||_2^2
\end{eqnarray} 
\vspace{0.1in}

\noindent where $x_{cloaked}$ and $x_{org}$ is a pair of cloaked and original
artworks. This optimization effectively forces $\Phi$ to extract the same
feature representation for cloaked and original artwork. To prevent the
extractor from collapsing (\eg output zero vectors for all inputs), we
regularized the training with the standard VAE reconstruction loss and train
the decoder $D$ at the same time. Given the high discrepancy between features
of cloaked and original artwork, this training process significantly modifies
the internals of $\Phi$ as well as the feature space. Thus, the mimic needs
to fine-tune the decoder $D$ and generator $G$ on the new robust feature
space. We assume the mimic trains $\Phi$ for $K$ steps on $K$ different pairs of
cloaked/original artwork, and then fine-tunes $D$ and $G$ until convergence.

\begin{figure*}
  \centering
  \includegraphics[width=1.0\linewidth]{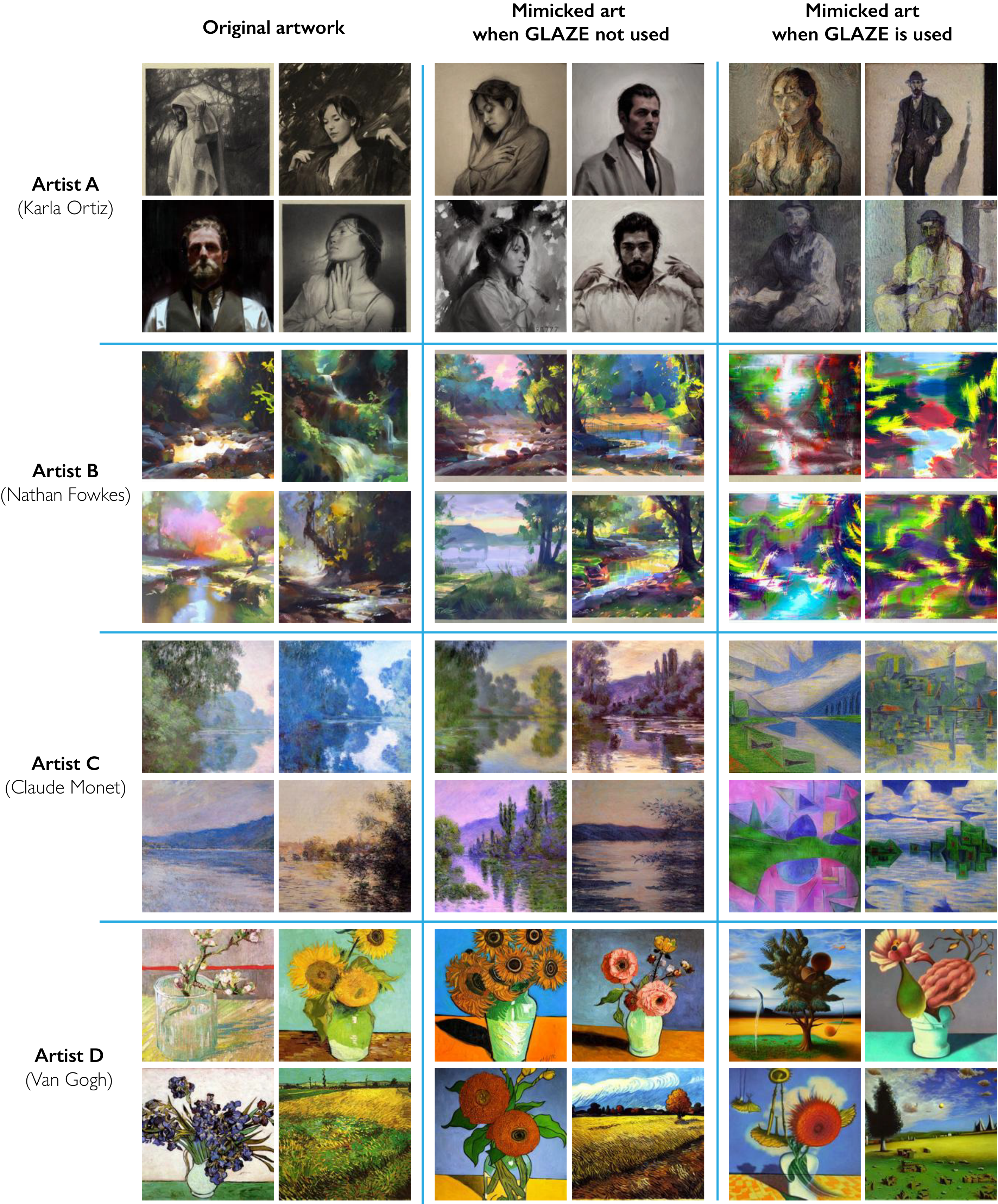}
  \caption{Additional example \system{} protection results for four artists. {\bf Columns 1-2}: artist's original artwork; {\bf column 3-4}: plagiarized artwork when artist does not use protection; {\bf column 5-6}: plagiarized artwork when artist uses cloaking protection with perturbation budget $p=0.05$. All mimicry attempts use SD-based models. }
  \label{fig:additional-pictures}
\end{figure*}

\end{document}